\PassOptionsToPackage{hyphens}{url}%
\PassOptionsToPackage{hidelinks}{hyperref}%
\documentclass[a4paper,man,natbib]{apa6}%

\usepackage[english]{babel}
\usepackage[utf8x]{inputenc}
\usepackage{amsmath}
\usepackage{graphicx}
\usepackage{comment}
\usepackage{hyperref}

\usepackage{fancyhdr}
\fancypagestyle{firstpage}{
	\fancyhf{} 
	\fancyhead[C]{%
		\footnotesize%
This is a preprint of the following chapter: Jonas Oppenlaender, The Cultivated Practices of Text-to-Image Generation, published
in Humane Autonomous Technology, ISBN 978-3-031-66527-1, edited by Rebekah Rousi, Catharina von Koskull, and Virpi Roto, 2024, Palgrave Macmillan reproduced with permission of Palgrave Macmillan.
The final authenticated version is available online at: \href{https://dx.doi.org/10.1007/978-3-031-66528-8}{https://dx.doi.org/10.1007/978-3-031-66528-8}
} 
}
\usepackage{xcolor}
\usepackage{graphicx}%
\usepackage{multirow}%
\usepackage{tabularx}%
\usepackage{framed}%
\newcommand\todo[1]{{\leavevmode\color{black}{#1}}}%
\newcommand\chatgpt[1]{{\leavevmode\color{black}{#1}}}%

\usepackage{url}%
\PassOptionsToPackage{hyphens}{url}%

\title{
The Cultivated Practices of Text-to-Image Generation
}%

\shorttitle{}

 \author{Jonas Oppenlaender}
 \affiliation{
 oppenlaenderj@acm.org}

\abstract{%
Humankind is entering a novel creative era in which anybody can synthesize digital information 
using generative artificial intelligence (AI).
Text-to-image generation, in particular, has become vastly popular and millions of practitioners produce AI-generated images and AI art online.
This chapter first gives an overview of the key developments that enabled a healthy co-creative online ecosystem around text-to-image generation to rapidly emerge, followed by a high-level description of key elements in this ecosystem.
A particular focus is placed on prompt engineering, a creative practice that has been embraced by the AI art community. 
It is then argued that the emerging co-creative ecosystem
constitutes an intelligent system on its own --- a system that both
supports human creativity, but also potentially entraps future generations and limits future development efforts in AI.
The chapter discusses the potential risks and dangers of cultivating this co-creative ecosystem, such as the bias inherent in today's training data, potential quality degradation in future image generation systems due to synthetic data becoming common place, and the potential long-term effects of text-to-image generation on people's imagination, ambitions, and development.
\\[\baselineskip]

\\[2\baselineskip]

}

\begin{document}
\thispagestyle{firstpage}
\maketitle

\section{Introduction}


Generative artificial intelligence (AI) has taken the world by storm.
Using deep generative models, anybody can conjure up digital information from short descriptive text prompts.
Text-to-image synthesis, in particular, has become a popular means for generating digital images \citep{2207.13038.pdf,VQGANCLIP}.
    Millions of people use generative systems and text-to-image services available online, such as Midjourney\footnote{https://www.midjourney.com}, Stable Diffusion \citep{latent-diffusion}, and DALL-E 2 \citep{dalle2}, both for professional and recreational uses.
With this powerful generative technology at our fingertips, humankind is ushering into a new era --- an era in which visual imagery no longer necessarily reflects the effort put into creating the imagery \citep{mindtrek-creativity}.

This chapter first gives an overview of the key technical developments that enabled a co-creative ecosystem around text-to-image generation to rapidly emerge and expand in 2021 and 2022.
This is followed by a high-level description of key elements in the ecosystem and its practices.
    A focus is placed on prompt engineering, a method and creative practice that has proven useful in a broad set of application areas, but has been particularly embraced by the
    community of text-to-image generation practitioners.
%
It is then argued that the creative online ecosystem constitutes an intelligent system on its own --- a system that both enables but also potentially 
limits the creative potential of future generations of humans and machine learning systems.
We discuss some potential risks and dangers of cultivating 
this co-creative ecosystem, such as the threat of bias due to Western ways of seeing encoded in training data, quality degradation due to synthetic data being used for training future generative systems, and the potential long-term effects of text-to-image generation on people's creativity, imagination, and development.
%

\section{Background on text-to-image generation}%
The history of computer-generated art and ``generative art'' \citep{galanter_generative.pdf,boden2009.pdf}
goes back to first experiments with AI \citep{whatisanimage.pdf}.
Looking back, the first attempts to synthesize images from text were humbling, but already showed great promise.
The synthetic images presented by \cite{1511.02793.pdf}, for instance, were tiny in size (e.g., a 32x32 pixel resolution image of a ``green school bus'').
Today, text-guided synthesis of images has made a giant leap towards becoming a mainstream phenomenon \citep{messy.pdf}.
Within less than a year, Midjourney's Discord community has grown to over 10 million users, making Midjourney the largest Discord community to date.\footnote{See https://discord.com/servers.}
Besides more powerful graphics processing units (GPUs), a few particular important inventions advanced the field of text-to-image generation.
This section gives a brief overview aiming to explain the recent technical developments that enabled and fueled the meteoric rise of 
text-to-image generation.


The invention of Generative Adversarial Networks (GANs) \citep{GAN} was a watershed moment in advancing image generation. GANs are a type of deep learning architecture consisting of two antagonistic parts: a generator and discriminator.
        During training, the generator presents the discriminator with synthetic images. The discriminator judges these images and the process is iteratively continued until the discriminator cannot tell the synthetic images apart from real images, such as the images in the training data.
Using a text-conditioned GAN architecture, \cite{1605.05396.pdf} pioneered the approach of synthesizing images from text.
The approach 
was extended in January 2021 with OpenAI's DALL-E \citep{dalle2}, a neural network trained on text-image pairs. 
DALL-E was able to synthesize images from text captions for a wide range of concepts expressible in natural language.
In parallel, OpenAI presented CLIP \citep{CLIP}, a contrastive language-vision model originally conceived for the task of classifying images. 
CLIP was trained on a large corpus of pairs of images and text scraped from the World Wide Web. Due to the large size of its training data, the CLIP model has learned a wide variety of visual concepts from natural language supervision.
This proved useful for tasks that visually associate language with images.
The CLIP model and its training corpus were, however, not released by OpenAI, which spurred efforts to replicate CLIP and its training data.\\
It was the release of the weights of CLIP in January 2021 that resulted in immense technical progress in the field of AI-generated imagery.
The CLIP weights found their first significant application in an image generation system called ``The Big Sleep'' by Ryan Murdoch \citep{bigsleepgithub,bigsleepiccc}.
    In Murdoch's architecture, the generator is a model called BigGAN and CLIP is used to guide the generation process with text.
This inspired Katherine Crowson to connect a more powerful neural network (VQGAN) with CLIP~\citep{VQGANCLIP}.
The VQGAN--CLIP architecture became very popular in 2021 and
instrumental to advancing the emerging field of text-to-image generation~\citep{VQGANCLIP}.
The source code of VQGAN--CLIP was available online, and many generative architectures for synthesizing digital images and artworks have since been developed based on the 
work by Murdoch and Crowson.
GANs were later superseded by diffusion-based systems~\citep{NEURIPS2021_49ad23d1}.
\chatgpt{%
Diffusion models are a class of machine learning models that learn through the introduction of incremental noise into the training data, with the objective of subsequently reversing the noising process and restoring the original image.
Once trained, these models are capable of utilizing the learned denoising methods to synthesize novel, noise-free images from random input.}%
\\
Today, practitioners can choose from a large variety of diffusion-based generative systems.
Some of these systems are available as open source, such as StableDiffusion \citep{latent-diffusion}, others are available as online services, such as Midjourney and DALL-E 2.
Due to the low barrier of entry and high ease of use, Colab notebooks contributed to a democratization of digital art production. Anybody can create digital images and artworks with text-to-image generation systems \citep{mindtrek-creativity},
which establishes parallels of the novel technology with photography.
\section{Text-to-image generation as the new photography}%
As a novel phenomenon and emerging technology,
text-to-image generation --- and generative AI in general --- can be compared to past disruptive and transformative technologies.
For instance, the invention of Gutenberg's printing press reduced the cost of printing, revolutionized the spread of knowledge, and had a profound impact on human development \citep{printingpress}.
Generative AI could have similar transformative effects on society.
This section briefly discusses the parallels between the invention of photography and text-to-image generation, followed by some current criticisms of text-to-image generation.

\subsection{Parallels between photography and text-to-image generation}%
When photography was invented, 
    critics argued against the new technology.
Photography's early critics saw the novel technology not as a new medium for creative expression, but as a direct threat to the livelihood of artists.
As \citeauthor{arts-07-00018-v3.pdf} points out,
``[m]any artists were dismissive of photography, and saw it as a threat to `real art'{''} \citep{arts-07-00018-v3.pdf}.
    For instance, upon seeing a demonstration of the daguerreotype technique, the painter Paul Delaroche declared: ``From today, painting is dead!''
\citep{arts-07-00018-v3.pdf}.
\chatgpt{%
Photography, as a mechanical way of capturing reality, was seen in direct competition to realism, an art style that aims to depict nature accurately and in great detail.
But over the years, photography has evolved into a medium for artistic expression, a medium that allows for the creation of unique images that were previously unimaginable.
}%
Ultimately, the invention of photography caused painters to innovate their craft, moving from realism to abstraction \citep{arts-07-00018-v3.pdf,3347092.pdf}.

Text-to-image generation, as a new artistic medium \citep{Dont_Fear_the_Artwork_of_the_Future_-_The_Atlantic.pdf},
\chatgpt{offers equally exciting new possibilities for artistic expression and creativity.}
The emerging technology has the potential to revolutionize the way we work creatively and may fundamentally impact our relationship to digital media.
    \chatgpt{In many ways, text-to-image generation is \textit{the new photography},
democratizing access to creative expression that was previously limited to highly talented creative individuals.}
Text-to-image generation, however, also poses many complex socio-technical challenges with no easy solution.
\chatgpt{%
While it is undeniable that AI algorithms can generate fascinating images, it is important to consider the potential criticisms
of this technology.
}%

\subsection{Criticisms about text-to-image generation}%
Today's concerns and criticisms about text-to-image generation resemble the ones that were being raised when photography was invented \citep{arts-07-00018-v3.pdf}.
This chapter summarizes some of the key concerns and criticisms about generative AI, and text-to-image generation in particular.
\todo{%
\chatgpt{As we continue to develop and explore generative AI, it will be crucial to address these concerns and ensure that the technology is used in a responsible and ethical way.}}%
\\
%
One of the main criticisms about text-to-image generation is that it threatens to automate and replace human cognitive and creative work.
    Generative AI is much faster and cheaper than human work,
    and it does not require much skill and effort to prompt generative systems \citep{mindtrek-creativity}.
    This could lead companies to stop hiring or contracting knowledge workers and creatives, such as illustrators, designers, and artists \citep{Stack.pdf,oppenlaender2023perceptions}.\\
%
%
Another concern brought forth against the technology of text-to-image generation is that it is built on top of centuries of human-made creative works.
%
%
Generative AI is data hungry \citep{LLMs.md.pdf}
Whether it is literature used for training large language models or images scraped from the Web, generative models are trained on vast datasets scraped from the World Wide Web.
This ``Web-scale'' data enables generative models to learn patterns 
in the data. 
Critics of the technology argue that there are in\-sur\-mount\-able legal issues concerning the use of the training data.
Proponents, on the other hand, argue that using scraped training data falls under the U.S. fair use doctrine \citep{scraping}.
The fair use doctrine has opened a loophole in which commercial organizations are funding development efforts in non-profit organizations, which are allowed to train models and scrape data under fair and academic use.
These commercial organizations essentially are using non-profits as a shield from litigation and have been accused of data laundering \citep{data-laundering} and deceptive trade practices \citep{getty_images_vs_stability_AI_delaware.pdf}.
%
%
For instance, Microsoft has received criticism for training its Co-Pilot system on GitHub \citep{copilotcritic}.
    GitHub is a large repository of software licensed under various terms, including copyleft and MIT licenses.
    These licenses mandate that any modifications and derivitatives of the software must be free or distributed under the same license, respectively.
    Co-Pilot cannot --- by design --- guarantee that the license is adhered to and, due to the opaqueness of the neural network, it is not possible to trace the generated information to its source.
Similar concerns were raised about MegaFace, a large database of facial images scraped from Flickr \citep{1705.00393.pdf}. The database was used by numerous commercial organizations to advance the state of facial recognition for various purposes, including surveillance \citep{MegaFace.pdf}.
Legal battles between generative AI firms and content creators,
such as the class action law suits against Microsoft and OpenAI \citep{Verge.pdf} and the lawsuit of Getty Images against Stability AI \citep{getty_images_vs_stability_AI_delaware.pdf}, are also battles about the future direction of the creative industries.\\
%
%
\chatgpt{The emergence of generative AI as a novel and impactful technology is expected to significantly disrupt the current status quo, creating frictions and challenges in relation to existing societal norms and policy frameworks}
\citep{Predictions}. 
Generative AI is a direct threat to some business models.
For instance, Google's search engine business may become affected if consumer preferences shift from search engines to query-answering language models.
Stackoverflow is another company that may be heavily impacted by generative AI. Answers provided by generative AI undermine Stackoverflow's community, which consists of human volunteers providing human-written answers to human-written questions. The company has banned generative AI from its websites \citep{stackoverflow}.
A similar ban was instated and legal action has been taken by the stock photography website Getty Images \citep{getty,getty-infringement,getty_images_vs_stability_AI_delaware.pdf}.
The above examples highlight the cross-sectional and complex impact of generative AI on a variety of businesses and entire industries.\\
%
%
Another criticism about generative AI is that it may violate people's privacy.
    For instance, diffusion models have been shown to memorize and replicate training data \citep{2301.13188.pdf,2212.03860.pdf}.
That means diffusion models could reproduce near-exact matches of instances found in the training data. 
This could not only have copyright implications, but also lead to the unintentional release 
of private data.
    For instance, LAION --- the large dataset used to train StableDiffusion --- was shown to contain medical images of patients included in the dataset without consent of the patient \citep{LAION-patientdata}.
Memorization is a fundamental issue in diffusion models and may even be necessary for generative models to generalize
    \citep{1906.05271.pdf}.\\
On a less technical level,
\chatgpt{another criticism about text-to-image generation is that it could erode human creativity and artistic expression.
Critics argue that by relying on algorithms and machine learning to generate images, we are losing touch with the human elements that make art so special. Critics worry that AI art may become a sterile and impersonal medium, devoid of the emotional depth and individuality that characterizes great human art \citep{oppenlaender2023perceptions}.
Another related concern is that AI art could lead to the homogenization of artistic styles and forms. Because generative models are trained on vast datasets of images, they tend to produce art that is similar to what has come before.
Midjourney\footnote{Version 3}, for instance, produces images with a recognizable style.
Current generative systems ``lack a concept of novelty regarding how their product differs from previously created ones'' \citep{ICCC-2022_15L_Zammit-et-al..pdf}.
This could lead to a situation where all AI art begins to look the same, with little variation or originality.\\
%
%
%
Another concern is that text-to-image generation could be used to create false or misleading images \citep{oppenlaender2023perceptions}. With the ability to generate highly realistic images, AI algorithms could be used to create fake photographs or other forms of visual media for the purpose of spreading misinformation. This could have serious implications for the reliability of information and the trustworthiness of digital media that we encounter online.}\\
Even proponents of text-to-image generation have voiced concerns about the technology.
Given the improvements in the latest versions of text-to-image models, some practitioners of AI art have complained that it is becoming ``too easy'' to conjure up images from text prompts \citep{mindtrek-creativity,arstechnica}.
This raises the interesting question about the optimal skill level for text-to-image generation and AI-generated art. If the skill level is too difficult, users will not be able to communicate their intent to the model. The result could be that users 
become frustrated. On the other hand, if the skill level is too easy, users will not have a strong sense of ownership and feel that the generated images do not reflect their intent and are merely retrieved from pre-made collection of images.
The current generative systems vary in this regard.
    StableDiffusion, for instance, requires more effort being put into writing prompts than Midjourney.
This effort put into writing prompts is part of the novel practice of prompt engineering which is discussed in the following section.%
%
%
%
%
%
%
%
%
%
\section{The creative practice of prompt engineering}%
%
Prompt engineering~\citep{chilton,taxonomy} 
    --- or prompting 
    for short ---
is an interaction pattern in which machine learning models are given text as inputs 
\citep{2005.14165.pdf}.
Prompt engineering is a paradigm shift in how machine learning models are adapted to various downstream tasks. Instead of retraining or fine-tuning the model, the model is prompted with context.
    In zero-shot prompting \citep{2205.11916.pdf}, the user directly prompts the generative model, whereas in few-shot prompting \citep{2005.14165.pdf}, the user first provides a few examples to provide context to the  model.
Zero-shot prompting has seen an ideal application ground in AI-generated art. In the context of text-to-image generation, prompt engineering means that ``carefully selected and composed sentences are used to achieve a certain visual style in the synthesized image''~\citep{2207.13038.pdf}.\\
%
%
%
%
Prompt engineering is not an exact engineering science as found in science, technology, engineering, and mathematics (STEM).
Rather, its origins are within the online community of practitioners of text-to-image generation and AI artists who practice prompt engineering to exercise their craft.
Prompt engineering is iterative and resembles a conversation with the text-to-image system.
A practitioner typically will type a prompt, observe the outcome, and adapt the prompt to improve the outcome.\\
%
%
%
The online community around text-to-image generation found that the aesthetic qualities and subjective attractiveness of images can be modified by adding certain keywords to prompts \citep{taxonomy}.
By adding such modifiers to an input prompt, one seeks to direct the text-to-image model to produce images in a certain style or with a certain quality.
Knowing what prompt modifiers work best for a given subject term is often the result of the practitioner's iterative and persistent experimentation \citep{juho,chilton}.
Community-based resources have been created as education materials about the practice of prompt engineering~\citep{mindtrek-creativity}.
Supported by these numerous guidelines and learning resources, anybody can materialize seemingly creative artifacts with generative AI, whether it's textual (e.g., poems and essays written with OpenAI's GPT language models) or visual (e.g., photographic images with Midjourney and StableDiffusion),
with little to no understanding of the underlying technologies.
As generative AI matures and gains more prominence in our daily lives, it will become more important for us to be able to communicate effectively with the AI without having to resort to technical jargon and complicated keywords as is currently the case with prompt engineering.
Critics of prompt engineering, therefore, point out its many limitations:%
\begin{itemize}%
\item%
    Some text-to-image generation models accept only a limited number of tokens (e.g., 75 tokens). Tokens are pieces of text that roughly correspond to words in the prompt.
    Unless one picks a subject well represented in the training data (e.g., Leonardo da Vinci's Mona Lisa), it is difficult to describe a subject in detail to the generative AI with only a limited number of tokens.%
\item%
    Even if there is an unlimited number of tokens to describe an imagined image, it is not humanly possible to describe a subject in every minute detail. In practice, the generative model will need to fill in the blanks. This, in turn, can lead the practitioner to either settle for an image that is ``good enough'' or abandon the pursuit of the originally envisioned image \citep{mindtrek-creativity}. 
    On the other hand, the mismatch between the imagined output and the generated image can also spark creativity in the human user
    \citep{ICCC-2022_16S_Epstein-et-al.-1.pdf}.
\item%
    The keywords that humans use to describe a subject may not correspond to the concepts that the neural network has learned during its training process.
    Certain tokens in the prompt could also have unintended side-effects, for instance, on the style of an image, and concepts can ``leak'' from the prompt into the image \citep{2210.10606.pdf}. It is, for example, not uncommon that parts of the prompts will surface in generated images as text, even if that was not intended by the user.
\item%
    Each time a generative model is updated, the practice of prompt engineering is heavily affected and practitioners have to relearn their craft \citep{pehype}.
\end{itemize}%

The latter is supported by a wealth of learning resources that have emerged online to support practitioners of text-to-image generation in learning the craft of prompt engineering.
Together with communities and tools and services, these resources are the main pillars of the co-creative ecosystem of text-to-image generation.
%




\section{The human-computer co-creative ecosystem of text-to-image generation}%
%
Prompt engineering is a practice embedded in the greater online creative ecosystem of text-to-image generation that consists of communities, learning resources, and tools and services \citep{mindtrek-creativity}.%
%
%
\subsubsection{Communities dedicated to text-to-image generation}%
Online communities have emerged around text-guided generative art.
\chatgpt{%
These communities provide a platform for individuals to share their creative prompts and works of art, or even collaborate on generating new images collectively \citep{014-iccc20.pdf}.
The Midjourney community has emerged as a cultural hub contributing to the proliferation of prompt engineering, due to providing an accessible service and having an ear for the needs of its community members. 
Further examples of online communities include /r/MediaSynthesis on Reddit, and numerous other communities hosted on Discord. The latter, in particular, has proven to be an effective tool for community formation due to its ability to facilitate the creation of chat-based online communities and the possibility for community members to directly interact with  the image generation systems via chat.}\\
%
%
Online communities act as a fertile ground for learning the skills necessary to use text-to-image generation systems creatively.
\chatgpt{The online communities of enthusiastic creators who share their prompts and practices serve as a rich environment for novice practitioners to gain knowledge from other members of the community in order to surmount the challenges of prompt engineering. However, the distributed and sequential nature of communication within these communities presents a challenge for practitioners seeking to attain specific learning objectives, such as the study of prompt modifiers. As a result, there is a growing trend among practitioners and online communities to establish dedicated learning resources that cater to the learning needs of novices.}%
%
%
\subsubsection{Dedicated learning resources}%

\chatgpt{%
Community members have created various learning resources related to prompt engineering, including resources aimed at specific learning objectives, such as style experimentation \citep{DiscoDiffusionArtiststudies,artstudies} and teaching prompt engineering (e.g., the DALL-E prompt book by \cite{dallepromptbook}). Some of these resources adopt a similar approach to the systematic experimentation conducted by \cite{chilton}, presenting results of their experiments in tabular format.
Examples of these resources include the artist studies by \cite{artstudies},
``MisterRuffian's Latent Artist and Modifier Encyclopedia'' \citep{MisterRuffian},
and the list of Disco Diffusion modifiers by \cite{disco-diffusion-modifiers}.
Hub pages are another type of resource that have been established to improve the discoverability of the rapidly growing field of text-to-image generation. These hub pages function as indexes, compiling links to Colab notebooks and improving accessibility to the quickly growing number of resources. Two examples of hub pages are Miranda's list of VQGAN-CLIP implementations\footnote{https://ljvmiranda921.github.io/notebook/2021/08/11/vqgan-list/} and pharmapsychotic's ``Tools and Resources for AI Art''.\footnote{https://pharmapsychotic.com/tools.html}}


\subsubsection{Dedicated tools and services}%
A growing number of interfaces, tools, and services are emerging to support practitioners in practicing text-to-image generation.
For instance, a wide variety of text-to-image generation systems are available as open source in executable notebooks.
    Google's Colaboratory\footnote{https://colab.research.google.com} (Colab), in particular, has proven instrumental to the early growth and popularity of text-to-image generation.
    Colab is an online service that allows anybody to execute Python-based code and machine learning models for free.
The growing ecosystem of tools and services assists in making text-to-image generation more accessible to non-technically mind\-ed practitioners. The ecosystem also acts as a catalyst that draws in new practitioners and advances the field as a whole.
It is the specific combination of people, technology, services, tools, and resources that formed a healthy co-creative ecosystem for the text-to-image art community to thrive.
\section{The risks and dangers of cultivating the co-creative ecosystem}%
%
As the generative AI revolution advances, generative AI is becoming infused into many software applications and creative tools.
    For instance, users of Adobe Photoshop can create images from text using extensions based on Stable Diffusion  \citep{photoshop2,photoshop1}, and
    Microsoft's Co-Pilot \citep{copilot}
    has become an indispensable tool for many software developers.
    Generative AI is also explored in the field of generative design \citep{computers-09-00080.pdf,DreamLens-CHI-Postrebuttal.pdf} and architecture \citep{paananen2023architecture}.
Given the emerging ubiquity of generative models, such as large language models \citep{2005.14165.pdf,1810.04805.pdf,language_models_are_unsupervised_multitask_learners.pdf,1910.10683.pdf} and other foundation-scale models \citep{foundationmodels},
it is foreseeable that even more creative work will be completed with support of generative AI in the future.\\
%
%
These generative models and, thus, the tools and applications built with them, are based on deep learning. The results of deep neural networks are difficult to interpret and understand by both laypeople and experts \citep{1606.03490.pdf,1802.07810.pdf}.
With the emerging ubiquity of generative technologies, we are at risk of creating ``systems of opaque systems'' ingrained with difficult to understand, potentially flawed, and biased logic.
This section discusses three potential dangers of cultivating the emerging co-creative ecosystem.%
%
%
%
\subsection{Bias in training data and generative models}%
%
Current approaches to training generative AI rely on vast datasets collected from the World Wide Web.
    For example, ``The Pile'' \citep{2101.00027.pdf} is a popular data source for training language models.
    LAION \citep{2210.08402.pdf,2111.02114.pdf} is a dataset based on CommonCrawl\footnote{https://commoncrawl.org},
    a large dataset released by a non-profit organization that periodically scrapes data from the World Wide Web.
        LAION-5B contains over 5~Billion text-image pairs, with text from the `alt' attributes of HTML image elements \citep{2210.08402.pdf}.
These big datasets are used for training multi-modal 
generative models.

It is known that data on the Web is biased \citep{p54-baeza-yates.pdf}.
    Web-based data may contain content that violates human preferences \citep{2302.08582.pdf}.
    Web-based datasets may also encode Western ways of seeing due to over-representation of certain viewpoints in the data, creating intersectional issues reflecting discrimination and privilege.
    For instance, the English language is over-represented in LAION-5B, with 2.3 Billion of the 5.85 Billion image-text pairs in the dataset being in English language, and the rest representing more than 100 other languages or texts that cannot be assigned to a language \citep{2210.08402.pdf}.

Generative models trained on this biased data may repeat and in some cases amplify undesirable biases, such as
%
    demographic 
    biases 
    \citep{jinfopoli_8_1_78.pdf,0654.pdf,3442188.3445922.pdf,2207.10245.pdf,3334480.3382791.pdf}.
    \cite{2110.01963.pdf} found that LAION 
    contains many troublesome and explicit images and text pairs, including ``rape, pornography, malign stereotypes, racist and ethnic slurs, and other extremely problematic content.''
    \cite{2305.13873.pdf} found that text-to-image generators create a substantial amount 
    of unsafe images, including sexually explicit, violent, disturbing, hateful, and political images, which could be used to spread hateful memes.
This problematic content may surface in downstream applications  and cause harm \citep{3490155.pdf}.
That is one reason why deep neural networks are difficult to audit 
\citep{model-written-evals.pdf} and harm is often only discovered during everyday use \citep{2105.02980.pdf}.

%
%

\subsection{A flood of synthetic data}
%
Synthetic images are being shared en masse on social media.
This flood of synthetic data \citep{messy.pdf} raises 
concerns that  synthetic imagery could taint the training data of future generations of text-to-image models 
and perpetuate the weaknesses of current text-to-image systems.
    For instance, current text-to-image models  struggle with human anatomy, in particular the accurate depiction of human hands.
Data quality has been shown to be equally important than the size of the training data \citep{2203.15556.pdf}.
Using low-quality images as training data could result in further degradation in the quality of future generative models.
    \cite{2305.17493.pdf} call this phenomenon the ``curse of recursion.''
    Models trained on data generated by prior generative models may degenerate and forget the underlying data distribution, leading to learned behavior with limited variance \citep{2305.17493.pdf}.
\\
%
%
Some proponents of generative AI argue that future generative models will learn to associate a new category of images with synthetic imagery, and that we can simply prompt the generative model not to produce images that look like they were generated by AI.
However, this is 
based on the assumption that we can find a way to prompt the generative model to avoid the specific class of AI-generated images, which may or may not be possible.
The challenge is in finding the right keywords to denote the class of quality-degraded images, as mentioned in the section on prompt engineering.
Even if there was a clear label for the new class of synthetic imagery, it could happen that this synthetic look is inseparably associated with other classes.\\
%
%
A possible solution to this problem could be invisible watermarks applied to AI-generated images.
This metadata would allow the generative models to correctly distinguish the class of AI-generated imagery from human creations.
Another solution could be technological progress in image generation.
By training generative AI on smaller and more tightly controlled datasets, many of the problems could be avoided.
This sparse learning \citep{2104.08378.pdf} would also better mimic how humans learn from sparse cues in their environment.

\subsection{Long-term effects on individuals and human culture}%
%
By cultivating generative systems and adopting them in our creative practices, we risk becoming dependent on them.
Final-year students who finished their studies in 2020 are  the last cohort of students to complete their education without the support of generative AI. Future generations will be born into a co-creative ecosystem of ubiquitous generative AI.
Generative AI, in this regard, is similar to the Internet which had a profound impact on our lives, especially on the generation of 
``GenZ'' who grew up without ever knowing a time without the Internet.
The Internet has affected our cognition \citep{Firth,shallows}.
It has reconfigured our society and brought great benefits, but it had also drawbacks and risks.
With generative AI and text-to-image generation, there could be 
unforeseen direct and indirect negative consequences on society and individuals.\\
%
%
For one, generative AI could negatively affect people's imagination and cognitive functioning in the long-term.
We've already seen the negative and lasting effect of habitual use of mobile phones on people's cognition, memory, and attention span \citep{fpsyg-08-00605.pdf}.
One could argue that habituated practices of image generation 
could in the long term also negatively affect people's career ambitions.
For instance, the widespread use of AI could discourage would-be artists from pursuing a creative career \citep{Stack.pdf}.
Habitual use of generative AI could also negatively affect people's cognitive abilities, with lasting effects on imagination and effects on child development.
Aphantasia refers to the inability to visualize mental images 
\citep{tgab035.pdf}.
This ``ability to create a quasi-perceptual visual picture in the mind's eye'' \citep{1-s2.0-S1053810021001690-main.pdf} is important for daydreaming, imagination, and creativity \citep{1-s2.0-S0010945220301404-main.pdf}.
\cite{1-s2.0-S1053810021001690-main.pdf} found that about 4\% of a population of about 1000 people had a weakness or inability to create mental imagery.
Generative AI could potentially contribute to a rise in the prevalence of aphantasia in the general population.\\
%
%
%
%
%
Another effect could be that we become accustomed to lowering our expectations and settling for second-best options.
As discussed in the section on prompt engineering, the results of text-to-image generation are often random and do not match the mental image that the person writing the prompt had in mind.
The retrieval of an exact image from the generative model's ``infinite index'' \citep{2212.07476.pdf} can be an arduous task. With each new generation, the practitioner's efforts to ``retrieve'' images from the generative model may become derailed. The generative system may, for instance, present the user with interesting results that do not reflect the initial prompt, but are worth pursuing further.
This repeated 
settling for good enough results could lead to long-term changes in our ambition 
and could contribute to cultivating a culture of prototyping.\\
Another effect on culture could be a change in communication patterns.
Generative AI could lead to a shift in how we communicate with each other.
For instance, intelligent agents could write and summarize e-mails for us, a feat that OpenAI's ChatGPT already accomplishes quite well today.
Generative AI could also more directly support our communication, for instance, in the form of real-time translation of spoken words in face-to-face communications \citep{3379495.pdf}.
Text-to-image generation could contribute to the proliferation of memes on the Internet and fuel a meme-driven culture.
AI-generated media may allow creators to express feelings that could not be expressed through words.\\
%
%
However, the proliferation of synthetic media could lead to some human-created media becoming harder to find on the Web, with knock-on effects on humanity's knowledge and culture.
With the flood of synthetic media, the long-tail of information on the Web expands and becomes more noisy, which in turn makes information in the long-tail more difficult to retrieve. 
\cite{2211.08411.pdf} found that large language models struggle to learn the long-tail of knowledge.
If interaction with large language models becomes our primary way of answering queries, as opposed to searching the Web, long-tail knowledge could be lost.
This presents unique challenges to augmented AI \citep{2302.07842.pdf}.
Augmented models are a class of generative models that are equipped with the ability to use tools and access external knowledge bases. Such augmented models can, for instance, query APIs, execute functions, and retrieve information from search engines.
Synthetic media could negatively affect the operation of augmented models due to truthful information becoming harder to retrieve in search engines.\\
%
%
%
%
Text-to-image generation could also fundamentally change our relation to visual media and how we appreciate art.
AI-generated media is becoming ubiquitous, and it is not clear if pervasive synthetic media will make us appreciate human-made art more or less.
If anybody can create digital artworks that look like they were created by a masterful painter, will we still appreciate real paintings, whether digital or on a physical canvas?
The proliferation of AI-generated media makes it also harder for aesthetic trends to stick among many meaningless short-lived fads \citep{corecore}.
The model of attractive quality by \cite{kano} posits that products can contain ``exciting'' factors that contribute to the appreciation of the product. Over time, exciting factors become expected.
    The iPhone's touchscreen-based user experience, for instance, was exciting when it was first introduced, but later became the expected standard in mobile phones.
Perhaps, once 
the novelty of AI-generated media wears off and text-to-image generation will turn from excitement to being expected.
Then, generative AI's true potential will emerge.
Generative AI has the potential to become part of the fundamental layer on which we base our future society.
%
%
%
\section{Conclusion}%
\label{sec:conclusion}%
%


\chatgpt{%
Text-to-image generation technology has emerged as an exciting new area of creative practice, drawing parallels with photography in terms of its ability to visualize the world around us. However, as with any new technology, there are also concerns. The co-creative ecosystem of text-to-image generation, which involves both human and computer participants, raises issues related to bias in training data, a potential flood of synthetic data, and the long-term impact on individuals and human culture.

Despite these challenges, the creative practice of prompt engineering has already produced remarkable results, thanks in part to a flourishing online ecosystem of dedicated communities, learning resources, and tools and services. While these resources offer great potential for creativity and innovation, they also come with risks. Therefore, it is crucial to carefully consider the ethical implications of cultivating a co-creative ecosystem and to take steps to mitigate any potential negative effects.
Text-to-image generation has the potential to revolutionize the way we visualize and create new works of art, but it is essential to approach this technology with caution and responsibility.
As we continue to explore the possibilities of this exciting new field, we must be vigilant in identifying and addressing potential risks and dangers to ensure that generative AI is aligned with human values and that the benefits of generative AI are realized for the benefit of all.
}%





\bibliography{paper}

\begin{thebibliography}{}

\bibitem [\protect \citeauthoryear {%
Alfaraj%
}{%
Alfaraj%
}{%
{\protect \APACyear {2022}}%
}]{%
photoshop1}
\APACinsertmetastar {%
photoshop1}%
\begin{APACrefauthors}%
Alfaraj, A.%
\end{APACrefauthors}%
\unskip\
\newblock
\APACrefYearMonthDay{2022}{}{}.
\newblock
\APACrefbtitle {{Auto {Photoshop} {StableDiffusion} Plugin}.} {{Auto
  {Photoshop} {StableDiffusion} Plugin}.}
\newblock
\begin{APACrefURL}
  \url{https://github.com/AbdullahAlfaraj/Auto-Photoshop-StableDiffusion-Plugin}
  \end{APACrefURL}
\PrintBackRefs{\CurrentBib}

\bibitem [\protect \citeauthoryear {%
Baeza-Yates%
}{%
Baeza-Yates%
}{%
{\protect \APACyear {2018}}%
}]{%
p54-baeza-yates.pdf}
\APACinsertmetastar {%
p54-baeza-yates.pdf}%
\begin{APACrefauthors}%
Baeza-Yates, R.%
\end{APACrefauthors}%
\unskip\
\newblock
\APACrefYearMonthDay{2018}{may}{}.
\newblock
{\BBOQ}\APACrefatitle {Bias on the Web} {Bias on the web}.{\BBCQ}
\newblock
\APACjournalVolNumPages{Commun. ACM}{61}{6}{54–61}.
\newblock
\begin{APACrefDOI} \doi{10.1145/3209581} \end{APACrefDOI}
\PrintBackRefs{\CurrentBib}

\bibitem [\protect \citeauthoryear {%
Baio%
}{%
Baio%
}{%
{\protect \APACyear {2022}}%
}]{%
data-laundering}
\APACinsertmetastar {%
data-laundering}%
\begin{APACrefauthors}%
Baio, A.%
\end{APACrefauthors}%
\unskip\
\newblock
\APACrefYearMonthDay{2022}{}{}.
\newblock
\APACrefbtitle {AI Data Laundering: How Academic and Nonprofit Researchers
  Shield Tech Companies from Accountability.} {Ai data laundering: How academic
  and nonprofit researchers shield tech companies from accountability.}
\newblock
\begin{APACrefURL}
  \url{https://waxy.org/2022/09/ai-data-laundering-how-academic-and-nonprofit-researchers-shield-tech-companies-from-accountability/}
  \end{APACrefURL}
\PrintBackRefs{\CurrentBib}

\bibitem [\protect \citeauthoryear {%
Bender%
, Gebru%
, McMillan-Major%
\BCBL {}\ \BBA {} Shmitchell%
}{%
Bender%
\ \protect \BOthers {.}}{%
{\protect \APACyear {2021}}%
}]{%
3442188.3445922.pdf}
\APACinsertmetastar {%
3442188.3445922.pdf}%
\begin{APACrefauthors}%
Bender, E\BPBI M.%
, Gebru, T.%
, McMillan-Major, A.%
\BCBL {}\ \BBA {} Shmitchell, S.%
\end{APACrefauthors}%
\unskip\
\newblock
\APACrefYearMonthDay{2021}{}{}.
\newblock
{\BBOQ}\APACrefatitle {On the Dangers of Stochastic Parrots: Can Language
  Models Be Too Big?} {On the dangers of stochastic parrots: Can language
  models be too big?}{\BBCQ}
\newblock
\BIn{} \APACrefbtitle {Proceedings of the 2021 ACM Conference on Fairness,
  Accountability, and Transparency} {Proceedings of the 2021 acm conference on
  fairness, accountability, and transparency}\ (\BPG~610–623).
\newblock
\APACaddressPublisher{New York, NY, USA}{Association for Computing Machinery}.
\newblock
\begin{APACrefDOI} \doi{10.1145/3442188.3445922} \end{APACrefDOI}
\PrintBackRefs{\CurrentBib}

\bibitem [\protect \citeauthoryear {%
Birhane%
, Prabhu%
\BCBL {}\ \BBA {} Kahembwe%
}{%
Birhane%
\ \protect \BOthers {.}}{%
{\protect \APACyear {2021}}%
}]{%
2110.01963.pdf}
\APACinsertmetastar {%
2110.01963.pdf}%
\begin{APACrefauthors}%
Birhane, A.%
, Prabhu, V\BPBI U.%
\BCBL {}\ \BBA {} Kahembwe, E.%
\end{APACrefauthors}%
\unskip\
\newblock
\APACrefYearMonthDay{2021}{}{}.
\newblock
\APACrefbtitle {Multimodal datasets: Misogyny, pornography, and malignant
  stereotypes.} {Multimodal datasets: Misogyny, pornography, and malignant
  stereotypes.}
\newblock
\APACaddressPublisher{}{arXiv}.
\newblock
\begin{APACrefDOI} \doi{10.48550/ARXIV.2110.01963} \end{APACrefDOI}
\PrintBackRefs{\CurrentBib}

\bibitem [\protect \citeauthoryear {%
Boden%
\ \BBA {} Edmonds%
}{%
Boden%
\ \BBA {} Edmonds%
}{%
{\protect \APACyear {2009}}%
}]{%
boden2009.pdf}
\APACinsertmetastar {%
boden2009.pdf}%
\begin{APACrefauthors}%
Boden, M\BPBI A.%
\BCBT {}\ \BBA {} Edmonds, E\BPBI A.%
\end{APACrefauthors}%
\unskip\
\newblock
\APACrefYearMonthDay{2009}{}{}.
\newblock
{\BBOQ}\APACrefatitle {What is generative art?} {What is generative
  art?}{\BBCQ}
\newblock
\APACjournalVolNumPages{Digital Creativity}{20}{1-2}{21-46}.
\newblock
\begin{APACrefDOI} \doi{10.1080/14626260902867915} \end{APACrefDOI}
\PrintBackRefs{\CurrentBib}

\bibitem [\protect \citeauthoryear {%
Bommasani%
\ \protect \BOthers {.}}{%
Bommasani%
\ \protect \BOthers {.}}{%
{\protect \APACyear {2021}}%
}]{%
foundationmodels}
\APACinsertmetastar {%
foundationmodels}%
\begin{APACrefauthors}%
Bommasani, R.%
, Hudson, D\BPBI A.%
, Adeli, E.%
, Altman, R.%
, Arora, S.%
, von Arx, S.%
\BDBL {}et al.%
\end{APACrefauthors}%
\unskip\
\newblock
\APACrefYearMonthDay{2021}{}{}.
\newblock
{\BBOQ}\APACrefatitle {On the Opportunities and Risks of Foundation Models} {On
  the opportunities and risks of foundation models}.{\BBCQ}
\newblock
\APACjournalVolNumPages{CoRR}{abs/2108.07258}{}{}.
\PrintBackRefs{\CurrentBib}

\bibitem [\protect \citeauthoryear {%
Brown%
\ \protect \BOthers {.}}{%
Brown%
\ \protect \BOthers {.}}{%
{\protect \APACyear {2020}}%
}]{%
2005.14165.pdf}
\APACinsertmetastar {%
2005.14165.pdf}%
\begin{APACrefauthors}%
Brown, T\BPBI B.%
, Mann, B.%
, Ryder, N.%
, Subbiah, M.%
, Kaplan, J.%
, Dhariwal, P.%
\BDBL {}Amodei, D.%
\end{APACrefauthors}%
\unskip\
\newblock
\APACrefYearMonthDay{2020}{}{}.
\newblock
\APACrefbtitle {Language Models are Few-Shot Learners.} {Language models are
  few-shot learners.}
\newblock
\APACaddressPublisher{}{arXiv}.
\newblock
\begin{APACrefDOI} \doi{10.48550/ARXIV.2005.14165} \end{APACrefDOI}
\PrintBackRefs{\CurrentBib}

\bibitem [\protect \citeauthoryear {%
Carlini%
\ \protect \BOthers {.}}{%
Carlini%
\ \protect \BOthers {.}}{%
{\protect \APACyear {2023}}%
}]{%
2301.13188.pdf}
\APACinsertmetastar {%
2301.13188.pdf}%
\begin{APACrefauthors}%
Carlini, N.%
, Hayes, J.%
, Nasr, M.%
, Jagielski, M.%
, Sehwag, V.%
, Tramèr, F.%
\BDBL {}Wallace, E.%
\end{APACrefauthors}%
\unskip\
\newblock
\APACrefYearMonthDay{2023}{}{}.
\newblock
\APACrefbtitle {Extracting Training Data from Diffusion Models.} {Extracting
  training data from diffusion models.}
\newblock
\APACaddressPublisher{}{arXiv}.
\newblock
\begin{APACrefDOI} \doi{10.48550/ARXIV.2301.13188} \end{APACrefDOI}
\PrintBackRefs{\CurrentBib}

\bibitem [\protect \citeauthoryear {%
Carr%
}{%
Carr%
}{%
{\protect \APACyear {2011}}%
}]{%
shallows}
\APACinsertmetastar {%
shallows}%
\begin{APACrefauthors}%
Carr, N.%
\end{APACrefauthors}%
\unskip\
\newblock
\APACrefYear{2011}.
\newblock
\APACrefbtitle {The Shallows: What the Internet Is Doing to Our Brains} {The
  shallows: What the internet is doing to our brains}.
\newblock
\APACaddressPublisher{New York, N.Y., USA}{W. W. Norton \& Company, Inc.}
\PrintBackRefs{\CurrentBib}

\bibitem [\protect \citeauthoryear {%
Cohen%
}{%
Cohen%
}{%
{\protect \APACyear {1979}}%
}]{%
whatisanimage.pdf}
\APACinsertmetastar {%
whatisanimage.pdf}%
\begin{APACrefauthors}%
Cohen, H.%
\end{APACrefauthors}%
\unskip\
\newblock
\APACrefYearMonthDay{1979}{}{}.
\newblock
\APACrefbtitle {What is an Image?} {What is an image?}
\PrintBackRefs{\CurrentBib}

\bibitem [\protect \citeauthoryear {%
Colton%
, Smith%
, Berns%
, Murdock%
\BCBL {}\ \BBA {} Cook%
}{%
Colton%
\ \protect \BOthers {.}}{%
{\protect \APACyear {2021}}%
}]{%
bigsleepiccc}
\APACinsertmetastar {%
bigsleepiccc}%
\begin{APACrefauthors}%
Colton, S.%
, Smith, A.%
, Berns, S.%
, Murdock, R.%
\BCBL {}\ \BBA {} Cook, M.%
\end{APACrefauthors}%
\unskip\
\newblock
\APACrefYearMonthDay{2021}{}{}.
\newblock
{\BBOQ}\APACrefatitle {Generative Search Engines: Initial Experiments}
  {Generative search engines: Initial experiments}.{\BBCQ}
\newblock
\BIn{} \APACrefbtitle {Proceedings of the 12th International Conference on
  Computational Creativity} {Proceedings of the 12th international conference
  on computational creativity}\ (\BPG~237-246).
\newblock
\APACaddressPublisher{}{Association for Computational Creativity}.
\PrintBackRefs{\CurrentBib}

\bibitem [\protect \citeauthoryear {%
Crowson%
\ \protect \BOthers {.}}{%
Crowson%
\ \protect \BOthers {.}}{%
{\protect \APACyear {2022}}%
}]{%
VQGANCLIP}
\APACinsertmetastar {%
VQGANCLIP}%
\begin{APACrefauthors}%
Crowson, K.%
, Biderman, S.%
, Kornis, D.%
, Stander, D.%
, Hallahan, E.%
, Castricato, L.%
\BCBL {}\ \BBA {} Raff, E.%
\end{APACrefauthors}%
\unskip\
\newblock
\APACrefYearMonthDay{2022}{}{}.
\newblock
\APACrefbtitle {{VQGAN-CLIP}: Open Domain Image Generation and Editing with
  Natural Language Guidance.} {{VQGAN-CLIP}: Open domain image generation and
  editing with natural language guidance.}
\newblock
\APACaddressPublisher{}{arXiv}.
\newblock
\begin{APACrefDOI} \doi{10.48550/ARXIV.2204.08583} \end{APACrefDOI}
\PrintBackRefs{\CurrentBib}

\bibitem [\protect \citeauthoryear {%
Dance%
, Ipser%
\BCBL {}\ \BBA {} Simner%
}{%
Dance%
\ \protect \BOthers {.}}{%
{\protect \APACyear {2022}}%
}]{%
1-s2.0-S1053810021001690-main.pdf}
\APACinsertmetastar {%
1-s2.0-S1053810021001690-main.pdf}%
\begin{APACrefauthors}%
Dance, C.%
, Ipser, A.%
\BCBL {}\ \BBA {} Simner, J.%
\end{APACrefauthors}%
\unskip\
\newblock
\APACrefYearMonthDay{2022}{}{}.
\newblock
{\BBOQ}\APACrefatitle {The prevalence of aphantasia (imagery weakness) in the
  general population} {The prevalence of aphantasia (imagery weakness) in the
  general population}.{\BBCQ}
\newblock
\APACjournalVolNumPages{Consciousness and Cognition}{97}{}{103243}.
\newblock
\begin{APACrefDOI} \doi{10.1016/j.concog.2021.103243} \end{APACrefDOI}
\PrintBackRefs{\CurrentBib}

\bibitem [\protect \citeauthoryear {%
Danks%
\ \BBA {} London%
}{%
Danks%
\ \BBA {} London%
}{%
{\protect \APACyear {2017}}%
}]{%
0654.pdf}
\APACinsertmetastar {%
0654.pdf}%
\begin{APACrefauthors}%
Danks, D.%
\BCBT {}\ \BBA {} London, A\BPBI J.%
\end{APACrefauthors}%
\unskip\
\newblock
\APACrefYearMonthDay{2017}{}{}.
\newblock
{\BBOQ}\APACrefatitle {Algorithmic Bias in Autonomous Systems} {Algorithmic
  bias in autonomous systems}.{\BBCQ}
\newblock
\BIn{} \APACrefbtitle {Proceedings of the Twenty-Sixth International Joint
  Conference on Artificial Intelligence, {IJCAI-17}} {Proceedings of the
  twenty-sixth international joint conference on artificial intelligence,
  {IJCAI-17}}\ (\BPGS\ 4691--4697).
\newblock
\begin{APACrefDOI} \doi{10.24963/ijcai.2017/654} \end{APACrefDOI}
\PrintBackRefs{\CurrentBib}

\bibitem [\protect \citeauthoryear {%
Deckers%
\ \protect \BOthers {.}}{%
Deckers%
\ \protect \BOthers {.}}{%
{\protect \APACyear {2023}}%
}]{%
2212.07476.pdf}
\APACinsertmetastar {%
2212.07476.pdf}%
\begin{APACrefauthors}%
Deckers, N.%
, Fröbe, M.%
, Kiesel, J.%
, Pandolfo, G.%
, Schröder, C.%
, Stein, B.%
\BCBL {}\ \BBA {} Potthast, M.%
\end{APACrefauthors}%
\unskip\
\newblock
\APACrefYearMonthDay{2023}{}{}.
\newblock
{\BBOQ}\APACrefatitle {The Infinite Index: Information Retrieval on Generative
  Text-To-Image Models} {The infinite index: Information retrieval on
  generative text-to-image models}.{\BBCQ}
\newblock
\BIn{} \APACrefbtitle {ACM SIGIR Conference On Human Information Interaction
  And Retrieval.} {Acm sigir conference on human information interaction and
  retrieval.}
\PrintBackRefs{\CurrentBib}

\bibitem [\protect \citeauthoryear {%
Devlin%
, Chang%
, Lee%
\BCBL {}\ \BBA {} Toutanova%
}{%
Devlin%
\ \protect \BOthers {.}}{%
{\protect \APACyear {2019}}%
}]{%
1810.04805.pdf}
\APACinsertmetastar {%
1810.04805.pdf}%
\begin{APACrefauthors}%
Devlin, J.%
, Chang, M\BHBI W.%
, Lee, K.%
\BCBL {}\ \BBA {} Toutanova, K.%
\end{APACrefauthors}%
\unskip\
\newblock
\APACrefYearMonthDay{2019}{{\APACmonth{06}}}{}.
\newblock
{\BBOQ}\APACrefatitle {{BERT}: Pre-training of Deep Bidirectional Transformers
  for Language Understanding} {{BERT}: Pre-training of deep bidirectional
  transformers for language understanding}.{\BBCQ}
\newblock
\BIn{} \APACrefbtitle {Proceedings of the 2019 Conference of the North
  {A}merican Chapter of the Association for Computational Linguistics: Human
  Language Technologies, Volume 1 (Long and Short Papers)} {Proceedings of the
  2019 conference of the north {A}merican chapter of the association for
  computational linguistics: Human language technologies, volume 1 (long and
  short papers)}\ (\BPGS\ 4171--4186).
\newblock
\APACaddressPublisher{Minneapolis, Minnesota}{Association for Computational
  Linguistics}.
\newblock
\begin{APACrefDOI} \doi{10.18653/v1/N19-1423} \end{APACrefDOI}
\PrintBackRefs{\CurrentBib}

\bibitem [\protect \citeauthoryear {%
Dhariwal%
\ \BBA {} Nichol%
}{%
Dhariwal%
\ \BBA {} Nichol%
}{%
{\protect \APACyear {2021}}%
}]{%
NEURIPS2021_49ad23d1}
\APACinsertmetastar {%
NEURIPS2021_49ad23d1}%
\begin{APACrefauthors}%
Dhariwal, P.%
\BCBT {}\ \BBA {} Nichol, A.%
\end{APACrefauthors}%
\unskip\
\newblock
\APACrefYearMonthDay{2021}{}{}.
\newblock
{\BBOQ}\APACrefatitle {Diffusion Models Beat {GANs} on Image Synthesis}
  {Diffusion models beat {GANs} on image synthesis}.{\BBCQ}
\newblock
\BIn{} M.~Ranzato, A.~Beygelzimer, Y.~Dauphin, P.~Liang\BCBL {}\ \BBA {} J\BPBI
  W.~Vaughan\ (\BEDS), \APACrefbtitle {Advances in Neural Information
  Processing Systems} {Advances in neural information processing systems}\
  (\BVOL~34, \BPGS\ 8780--8794).
\newblock
\APACaddressPublisher{}{Curran Associates, Inc.}
\newblock
\begin{APACrefURL}
  \url{https://proceedings.neurips.cc/paper/2021/file/49ad23d1ec9fa4bd8d77d02681df5cfa-Paper.pdf}
  \end{APACrefURL}
\PrintBackRefs{\CurrentBib}

\bibitem [\protect \citeauthoryear {%
Durant%
}{%
Durant%
}{%
{\protect \APACyear {2021}}%
}]{%
artstudies}
\APACinsertmetastar {%
artstudies}%
\begin{APACrefauthors}%
Durant, R.%
\end{APACrefauthors}%
\unskip\
\newblock
\APACrefYearMonthDay{2021}{}{}.
\newblock
\APACrefbtitle {Artist Studies by @remi\_durant.} {Artist studies by
  @remi\_durant.}
\newblock
\begin{APACrefURL} \url{https://remidurant.com/artists/} \end{APACrefURL}
\PrintBackRefs{\CurrentBib}

\bibitem [\protect \citeauthoryear {%
Edwards%
}{%
Edwards%
}{%
{\protect \APACyear {2022}}%
{\protect \APACexlab {{\protect \BCnt {1}}}}}]{%
LAION-patientdata}
\APACinsertmetastar {%
LAION-patientdata}%
\begin{APACrefauthors}%
Edwards, B.%
\end{APACrefauthors}%
\unskip\
\newblock
\APACrefYearMonthDay{2022{\protect \BCnt {1}}}{}{}.
\newblock
\APACrefbtitle {Artist finds private medical record photos in popular {AI}
  training data set.} {Artist finds private medical record photos in popular
  {AI} training data set.}
\newblock
\begin{APACrefURL}
  \url{https://arstechnica.com/information-technology/2022/09/artist-finds-private-medical-record-photos-in-popular-ai-training-data-set}
  \end{APACrefURL}
\PrintBackRefs{\CurrentBib}

\bibitem [\protect \citeauthoryear {%
Edwards%
}{%
Edwards%
}{%
{\protect \APACyear {2022}}%
{\protect \APACexlab {{\protect \BCnt {2}}}}}]{%
arstechnica}
\APACinsertmetastar {%
arstechnica}%
\begin{APACrefauthors}%
Edwards, B.%
\end{APACrefauthors}%
\unskip\
\newblock
\APACrefYearMonthDay{2022{\protect \BCnt {2}}}{}{}.
\newblock
\APACrefbtitle {``{Too} easy'' -- {Midjourney} tests dramatic new version of
  its {AI} image generator.} {``{Too} easy'' -- {Midjourney} tests dramatic new
  version of its {AI} image generator.}
\newblock
\begin{APACrefURL}
  \url{https://arstechnica.com/information-technology/2022/11/midjourney-turns-heads-with-quality-leap-in-new-ai-image-generator-version/}
  \end{APACrefURL}
\PrintBackRefs{\CurrentBib}

\bibitem [\protect \citeauthoryear {%
Eisenstein%
}{%
Eisenstein%
}{%
{\protect \APACyear {1980}}%
}]{%
printingpress}
\APACinsertmetastar {%
printingpress}%
\begin{APACrefauthors}%
Eisenstein, E\BPBI L.%
\end{APACrefauthors}%
\unskip\
\newblock
\APACrefYear{1980}.
\newblock
\APACrefbtitle {The Printing Press as an Agent of Change} {The printing press
  as an agent of change}.
\newblock
\APACaddressPublisher{}{Cambridge University Press}.
\PrintBackRefs{\CurrentBib}

\bibitem [\protect \citeauthoryear {%
Epstein%
, Schroeder%
\BCBL {}\ \BBA {} Newman%
}{%
Epstein%
\ \protect \BOthers {.}}{%
{\protect \APACyear {2022}}%
}]{%
ICCC-2022_16S_Epstein-et-al.-1.pdf}
\APACinsertmetastar {%
ICCC-2022_16S_Epstein-et-al.-1.pdf}%
\begin{APACrefauthors}%
Epstein, Z.%
, Schroeder, H.%
\BCBL {}\ \BBA {} Newman, D.%
\end{APACrefauthors}%
\unskip\
\newblock
\APACrefYearMonthDay{2022}{}{}.
\newblock
{\BBOQ}\APACrefatitle {When happy accidents spark creativity: Bringing
  collaborative speculation to life with generative AI} {When happy accidents
  spark creativity: Bringing collaborative speculation to life with generative
  ai}.{\BBCQ}
\newblock
\BIn{} \APACrefbtitle {nternational Conference on Computational Creativity.}
  {nternational conference on computational creativity.}
\newblock
\APACaddressPublisher{}{arXiv}.
\newblock
\begin{APACrefDOI} \doi{10.48550/ARXIV.2206.00533} \end{APACrefDOI}
\PrintBackRefs{\CurrentBib}

\bibitem [\protect \citeauthoryear {%
Feldman%
}{%
Feldman%
}{%
{\protect \APACyear {2019}}%
}]{%
1906.05271.pdf}
\APACinsertmetastar {%
1906.05271.pdf}%
\begin{APACrefauthors}%
Feldman, V.%
\end{APACrefauthors}%
\unskip\
\newblock
\APACrefYearMonthDay{2019}{}{}.
\newblock
\APACrefbtitle {Does Learning Require Memorization? {A} Short Tale about a Long
  Tail.} {Does learning require memorization? {A} short tale about a long
  tail.}
\newblock
\APACaddressPublisher{}{arXiv}.
\newblock
\begin{APACrefDOI} \doi{10.48550/ARXIV.1906.05271} \end{APACrefDOI}
\PrintBackRefs{\CurrentBib}

\bibitem [\protect \citeauthoryear {%
Firth%
\ \protect \BOthers {.}}{%
Firth%
\ \protect \BOthers {.}}{%
{\protect \APACyear {2019}}%
}]{%
Firth}
\APACinsertmetastar {%
Firth}%
\begin{APACrefauthors}%
Firth, J.%
, Torous, J.%
, Stubbs, B.%
, Firth, J\BPBI A.%
, Steiner, G\BPBI Z.%
, Smith, L.%
\BDBL {}others%
\end{APACrefauthors}%
\unskip\
\newblock
\APACrefYearMonthDay{2019}{}{}.
\newblock
{\BBOQ}\APACrefatitle {The ``online brain'': How the {Internet} may be changing
  our cognition} {The ``online brain'': How the {Internet} may be changing our
  cognition}.{\BBCQ}
\newblock
\APACjournalVolNumPages{World Psychiatry}{18}{2}{119--129}.
\PrintBackRefs{\CurrentBib}

\bibitem [\protect \citeauthoryear {%
Gabha%
}{%
Gabha%
}{%
{\protect \APACyear {2022}}%
{\protect \APACexlab {{\protect \BCnt {1}}}}}]{%
DiscoDiffusionArtiststudies}
\APACinsertmetastar {%
DiscoDiffusionArtiststudies}%
\begin{APACrefauthors}%
Gabha, H.%
\end{APACrefauthors}%
\unskip\
\newblock
\APACrefYearMonthDay{2022{\protect \BCnt {1}}}{}{}.
\newblock
\APACrefbtitle {Disco {Diffusion} 70+ Artist Studies.} {Disco {Diffusion} 70+
  artist studies.}
\newblock
\begin{APACrefURL}
  \url{https://weirdwonderfulai.art/resources/disco-diffusion-70-plus-artist-studies/}
  \end{APACrefURL}
\PrintBackRefs{\CurrentBib}

\bibitem [\protect \citeauthoryear {%
Gabha%
}{%
Gabha%
}{%
{\protect \APACyear {2022}}%
{\protect \APACexlab {{\protect \BCnt {2}}}}}]{%
disco-diffusion-modifiers}
\APACinsertmetastar {%
disco-diffusion-modifiers}%
\begin{APACrefauthors}%
Gabha, H.%
\end{APACrefauthors}%
\unskip\
\newblock
\APACrefYearMonthDay{2022{\protect \BCnt {2}}}{}{}.
\newblock
\APACrefbtitle {Disco {Diffusion} Modifiers.} {Disco {Diffusion} modifiers.}
\newblock
\begin{APACrefURL}
  \url{https://weirdwonderfulai.art/resources/disco-diffusion-modifiers/}
  \end{APACrefURL}
\PrintBackRefs{\CurrentBib}

\bibitem [\protect \citeauthoryear {%
Galanter%
}{%
Galanter%
}{%
{\protect \APACyear {2016}}%
}]{%
galanter_generative.pdf}
\APACinsertmetastar {%
galanter_generative.pdf}%
\begin{APACrefauthors}%
Galanter, P.%
\end{APACrefauthors}%
\unskip\
\newblock
\APACrefYearMonthDay{2016}{}{}.
\newblock
{\BBOQ}\APACrefatitle {Generative Art Theory} {Generative art theory}.{\BBCQ}
\newblock
\BIn{} \APACrefbtitle {A Companion to Digital Art} {A companion to digital
  art}\ (\BPG~146-180).
\newblock
\APACaddressPublisher{Chichester, West Sussex, UK}{John Wiley \& Sons, Ltd}.
\newblock
\begin{APACrefDOI} \doi{10.1002/9781118475249.ch5} \end{APACrefDOI}
\PrintBackRefs{\CurrentBib}

\bibitem [\protect \citeauthoryear {%
Gao%
\ \protect \BOthers {.}}{%
Gao%
\ \protect \BOthers {.}}{%
{\protect \APACyear {2020}}%
}]{%
2101.00027.pdf}
\APACinsertmetastar {%
2101.00027.pdf}%
\begin{APACrefauthors}%
Gao, L.%
, Biderman, S.%
, Black, S.%
, Golding, L.%
, Hoppe, T.%
, Foster, C.%
\BDBL {}Leahy, C.%
\end{APACrefauthors}%
\unskip\
\newblock
\APACrefYearMonthDay{2020}{}{}.
\newblock
\APACrefbtitle {The {Pile}: {An} {800GB} Dataset of Diverse Text for Language
  Modeling.} {The {Pile}: {An} {800GB} dataset of diverse text for language
  modeling.}
\newblock
\begin{APACrefDOI} \doi{10.48550/ARXIV.2101.00027} \end{APACrefDOI}
\PrintBackRefs{\CurrentBib}

\bibitem [\protect \citeauthoryear {%
\APACcitebtitle {Getty {Images} ({US}), {Inc}. v. {Stability} {AI}, {Inc}.}}{%
\APACcitebtitle {Getty {Images} ({US}), {Inc}. v. {Stability} {AI}, {Inc}.}}{%
{\protect \APACyear {2023}}%
}]{%
getty_images_vs_stability_AI_delaware.pdf}
\APACinsertmetastar {%
getty_images_vs_stability_AI_delaware.pdf}%
\APACrefbtitle {Getty {Images} ({US}), {Inc}. v. {Stability} {AI}, {Inc}.}
  {Getty {Images} ({US}), {Inc}. v. {Stability} {AI}, {Inc}.}
\newblock
\APACrefYearMonthDay{2023}{}{}.
\newblock
\begin{APACrefURL}
  \url{https://fingfx.thomsonreuters.com/gfx/legaldocs/byvrlkmwnve/GETTY\%20IMAGES\%20AI\%20LAWSUIT\%20complaint.pdf}
  \end{APACrefURL}
\newblock
\APACrefnote{Case 1:23-cv-00135-UNA, Document 1}
\PrintBackRefs{\CurrentBib}

\bibitem [\protect \citeauthoryear {%
{GitHub Inc.}%
}{%
{GitHub Inc.}%
}{%
{\protect \APACyear {2021}}%
}]{%
copilot}
\APACinsertmetastar {%
copilot}%
\begin{APACrefauthors}%
{GitHub Inc.}%
\end{APACrefauthors}%
\unskip\
\newblock
\APACrefYearMonthDay{2021}{}{}.
\newblock
\APACrefbtitle {{GitHub} {Copilot} -- Your {AI} pair programmer.} {{GitHub}
  {Copilot} -- your {AI} pair programmer.}
\newblock
\begin{APACrefURL} \url{https://copilot.github.com} \end{APACrefURL}
\PrintBackRefs{\CurrentBib}

\bibitem [\protect \citeauthoryear {%
Goldberg%
}{%
Goldberg%
}{%
{\protect \APACyear {2023}}%
}]{%
LLMs.md.pdf}
\APACinsertmetastar {%
LLMs.md.pdf}%
\begin{APACrefauthors}%
Goldberg, Y.%
\end{APACrefauthors}%
\unskip\
\newblock
\APACrefYearMonthDay{2023}{}{}.
\newblock
\APACrefbtitle {Some remarks on Large Language Models.} {Some remarks on large
  language models.}
\newblock
\begin{APACrefURL}
  \url{https://gist.github.com/yoavg/59d174608e92e845c8994ac2e234c8a9}
  \end{APACrefURL}
\PrintBackRefs{\CurrentBib}

\bibitem [\protect \citeauthoryear {%
Goodfellow%
\ \protect \BOthers {.}}{%
Goodfellow%
\ \protect \BOthers {.}}{%
{\protect \APACyear {2014}}%
}]{%
GAN}
\APACinsertmetastar {%
GAN}%
\begin{APACrefauthors}%
Goodfellow, I.%
, Pouget-Abadie, J.%
, Mirza, M.%
, Xu, B.%
, Warde-Farley, D.%
, Ozair, S.%
\BDBL {}Bengio, Y.%
\end{APACrefauthors}%
\unskip\
\newblock
\APACrefYearMonthDay{2014}{}{}.
\newblock
{\BBOQ}\APACrefatitle {Generative Adversarial Nets} {Generative adversarial
  nets}.{\BBCQ}
\newblock
\BIn{} Z.~Ghahramani, M.~Welling, C.~Cortes, N.~Lawrence\BCBL {}\ \BBA {}
  K\BPBI Q.~Weinberger\ (\BEDS), \APACrefbtitle {Advances in Neural Information
  Processing Systems} {Advances in neural information processing systems}\
  (\BVOL~27).
\newblock
\APACaddressPublisher{}{Curran Associates, Inc.}
\PrintBackRefs{\CurrentBib}

\bibitem [\protect \citeauthoryear {%
Harvey%
\ \BBA {} LaPlace%
}{%
Harvey%
\ \BBA {} LaPlace%
}{%
{\protect \APACyear {2021}}%
}]{%
MegaFace.pdf}
\APACinsertmetastar {%
MegaFace.pdf}%
\begin{APACrefauthors}%
Harvey, A.%
\BCBT {}\ \BBA {} LaPlace, J.%
\end{APACrefauthors}%
\unskip\
\newblock
\APACrefYearMonthDay{2021}{}{}.
\newblock
\APACrefbtitle {Megaface.} {Megaface.}
\newblock
\begin{APACrefURL} \url{https://exposing.ai/megaface/} \end{APACrefURL}
\PrintBackRefs{\CurrentBib}

\bibitem [\protect \citeauthoryear {%
Hertzmann%
}{%
Hertzmann%
}{%
{\protect \APACyear {2018}}%
}]{%
arts-07-00018-v3.pdf}
\APACinsertmetastar {%
arts-07-00018-v3.pdf}%
\begin{APACrefauthors}%
Hertzmann, A.%
\end{APACrefauthors}%
\unskip\
\newblock
\APACrefYearMonthDay{2018}{}{}.
\newblock
{\BBOQ}\APACrefatitle {Can Computers Create Art?} {Can computers create
  art?}{\BBCQ}
\newblock
\APACjournalVolNumPages{Arts}{7}{2}{}.
\newblock
\begin{APACrefDOI} \doi{10.3390/arts7020018} \end{APACrefDOI}
\PrintBackRefs{\CurrentBib}

\bibitem [\protect \citeauthoryear {%
Hertzmann%
}{%
Hertzmann%
}{%
{\protect \APACyear {2020}}%
}]{%
3347092.pdf}
\APACinsertmetastar {%
3347092.pdf}%
\begin{APACrefauthors}%
Hertzmann, A.%
\end{APACrefauthors}%
\unskip\
\newblock
\APACrefYearMonthDay{2020}{apr}{}.
\newblock
{\BBOQ}\APACrefatitle {Computers Do Not Make Art, People Do} {Computers do not
  make art, people do}.{\BBCQ}
\newblock
\APACjournalVolNumPages{Commun. ACM}{63}{5}{45–48}.
\newblock
\begin{APACrefDOI} \doi{10.1145/3347092} \end{APACrefDOI}
\PrintBackRefs{\CurrentBib}

\bibitem [\protect \citeauthoryear {%
\APACcitebtitle {{HiQ} {Labs}, {Inc.} v. {LinkedIn} CORPORATION}}{%
\APACcitebtitle {{HiQ} {Labs}, {Inc.} v. {LinkedIn} CORPORATION}}{%
{\protect \APACyear {2022}}%
}]{%
scraping}
\APACinsertmetastar {%
scraping}%
\APACrefbtitle {{HiQ} {Labs}, {Inc.} v. {LinkedIn} CORPORATION.} {{HiQ} {Labs},
  {Inc.} v. {LinkedIn} corporation.}
\newblock
\APACrefYearMonthDay{2022}{}{}.
\newblock
\begin{APACrefURL}
  \url{https://cdn.ca9.uscourts.gov/datastore/opinions/2022/04/18/17-16783.pdf}
  \end{APACrefURL}
\newblock
\APACrefnote{No. 17-16783}
\PrintBackRefs{\CurrentBib}

\bibitem [\protect \citeauthoryear {%
Hoffmann%
\ \protect \BOthers {.}}{%
Hoffmann%
\ \protect \BOthers {.}}{%
{\protect \APACyear {2022}}%
}]{%
2203.15556.pdf}
\APACinsertmetastar {%
2203.15556.pdf}%
\begin{APACrefauthors}%
Hoffmann, J.%
, Borgeaud, S.%
, Mensch, A.%
, Buchatskaya, E.%
, Cai, T.%
, Rutherford, E.%
\BDBL {}Sifre, L.%
\end{APACrefauthors}%
\unskip\
\newblock
\APACrefYearMonthDay{2022}{}{}.
\newblock
\APACrefbtitle {Training Compute-Optimal Large Language Models.} {Training
  compute-optimal large language models.}
\newblock
\APACaddressPublisher{}{arXiv}.
\newblock
\begin{APACrefDOI} \doi{10.48550/ARXIV.2203.15556} \end{APACrefDOI}
\PrintBackRefs{\CurrentBib}

\bibitem [\protect \citeauthoryear {%
Kandpal%
, Deng%
, Roberts%
, Wallace%
\BCBL {}\ \BBA {} Raffel%
}{%
Kandpal%
\ \protect \BOthers {.}}{%
{\protect \APACyear {2022}}%
}]{%
2211.08411.pdf}
\APACinsertmetastar {%
2211.08411.pdf}%
\begin{APACrefauthors}%
Kandpal, N.%
, Deng, H.%
, Roberts, A.%
, Wallace, E.%
\BCBL {}\ \BBA {} Raffel, C.%
\end{APACrefauthors}%
\unskip\
\newblock
\APACrefYearMonthDay{2022}{}{}.
\newblock
\APACrefbtitle {Large Language Models Struggle to Learn Long-Tail Knowledge.}
  {Large language models struggle to learn long-tail knowledge.}
\newblock
\APACaddressPublisher{}{arXiv}.
\newblock
\begin{APACrefDOI} \doi{10.48550/ARXIV.2211.08411} \end{APACrefDOI}
\PrintBackRefs{\CurrentBib}

\bibitem [\protect \citeauthoryear {%
Kano%
, Seraku%
, Takahashi%
\BCBL {}\ \BBA {} Tsuji%
}{%
Kano%
\ \protect \BOthers {.}}{%
{\protect \APACyear {1984}}%
}]{%
kano}
\APACinsertmetastar {%
kano}%
\begin{APACrefauthors}%
Kano, N.%
, Seraku, N.%
, Takahashi, F.%
\BCBL {}\ \BBA {} Tsuji, S\BHBI i.%
\end{APACrefauthors}%
\unskip\
\newblock
\APACrefYearMonthDay{1984}{}{}.
\newblock
{\BBOQ}\APACrefatitle {Attractive Quality and Must-Be Quality} {Attractive
  quality and must-be quality}.{\BBCQ}
\newblock
\APACjournalVolNumPages{Journal of the Japanese Society for Quality
  Control}{14}{2}{147-156}.
\PrintBackRefs{\CurrentBib}

\bibitem [\protect \citeauthoryear {%
Kantosalo%
\ \BBA {} Takala%
}{%
Kantosalo%
\ \BBA {} Takala%
}{%
{\protect \APACyear {2020}}%
}]{%
014-iccc20.pdf}
\APACinsertmetastar {%
014-iccc20.pdf}%
\begin{APACrefauthors}%
Kantosalo, A.%
\BCBT {}\ \BBA {} Takala, T.%
\end{APACrefauthors}%
\unskip\
\newblock
\APACrefYearMonthDay{2020}{{\APACmonth{09}}}{7}.
\newblock
{\BBOQ}\APACrefatitle {Five {C}{\textquoteright}s for Human–Computer
  Co-Creativity: An Update on Classical Creativity Perspectives} {Five
  {C}{\textquoteright}s for human–computer co-creativity: An update on
  classical creativity perspectives}.{\BBCQ}
\newblock
\BIn{} A.~Cardoso, P.~Machado, T.~Veale\BCBL {}\ \BBA {} J.~Cunha\ (\BEDS),
  \APACrefbtitle {Proceedings of the 11th International Conference on
  Computational Creativity} {Proceedings of the 11th international conference
  on computational creativity}\ (\BPGS\ 17--24).
\newblock
\APACaddressPublisher{Portugal}{Association for Computational Creativity}.
\newblock
\APACrefnote{International Conference on Computational Creativity, ICCC ;
  Conference date: 07-09-2020 Through 11-09-2020}
\PrintBackRefs{\CurrentBib}

\bibitem [\protect \citeauthoryear {%
Kim%
}{%
Kim%
}{%
{\protect \APACyear {2022}}%
}]{%
juho}
\APACinsertmetastar {%
juho}%
\begin{APACrefauthors}%
Kim, J.%
\end{APACrefauthors}%
\unskip\
\newblock
\APACrefYearMonthDay{2022}{}{}.
\newblock
{\BBOQ}\APACrefatitle {Keynote on Interaction-Centric {AI}} {Keynote on
  interaction-centric {AI}}.{\BBCQ}
\newblock
\BIn{} \APACrefbtitle {{NeurIPS} 2022.} {{NeurIPS} 2022.}
\newblock
\begin{APACrefURL} \url{https://slideslive.com/38996064/interactioncentric-ai}
  \end{APACrefURL}
\PrintBackRefs{\CurrentBib}

\bibitem [\protect \citeauthoryear {%
Kirkpatrick%
}{%
Kirkpatrick%
}{%
{\protect \APACyear {2020}}%
}]{%
3379495.pdf}
\APACinsertmetastar {%
3379495.pdf}%
\begin{APACrefauthors}%
Kirkpatrick, K.%
\end{APACrefauthors}%
\unskip\
\newblock
\APACrefYearMonthDay{2020}{feb}{}.
\newblock
{\BBOQ}\APACrefatitle {Across the Language Barrier} {Across the language
  barrier}.{\BBCQ}
\newblock
\APACjournalVolNumPages{Commun. ACM}{63}{3}{15–17}.
\newblock
\begin{APACrefDOI} \doi{10.1145/3379495} \end{APACrefDOI}
\PrintBackRefs{\CurrentBib}

\bibitem [\protect \citeauthoryear {%
Kojima%
, Gu%
, Reid%
, Matsuo%
\BCBL {}\ \BBA {} Iwasawa%
}{%
Kojima%
\ \protect \BOthers {.}}{%
{\protect \APACyear {2022}}%
}]{%
2205.11916.pdf}
\APACinsertmetastar {%
2205.11916.pdf}%
\begin{APACrefauthors}%
Kojima, T.%
, Gu, S\BPBI S.%
, Reid, M.%
, Matsuo, Y.%
\BCBL {}\ \BBA {} Iwasawa, Y.%
\end{APACrefauthors}%
\unskip\
\newblock
\APACrefYearMonthDay{2022}{}{}.
\newblock
\APACrefbtitle {Large Language Models are Zero-Shot Reasoners.} {Large language
  models are zero-shot reasoners.}
\newblock
\APACaddressPublisher{}{arXiv}.
\newblock
\begin{APACrefDOI} \doi{10.48550/ARXIV.2205.11916} \end{APACrefDOI}
\PrintBackRefs{\CurrentBib}

\bibitem [\protect \citeauthoryear {%
Korbak%
\ \protect \BOthers {.}}{%
Korbak%
\ \protect \BOthers {.}}{%
{\protect \APACyear {2023}}%
}]{%
2302.08582.pdf}
\APACinsertmetastar {%
2302.08582.pdf}%
\begin{APACrefauthors}%
Korbak, T.%
, Shi, K.%
, Chen, A.%
, Bhalerao, R.%
, Buckley, C\BPBI L.%
, Phang, J.%
\BDBL {}Perez, E.%
\end{APACrefauthors}%
\unskip\
\newblock
\APACrefYearMonthDay{2023}{}{}.
\newblock
\APACrefbtitle {Pretraining Language Models with Human Preferences.}
  {Pretraining language models with human preferences.}
\newblock
\APACaddressPublisher{}{arXiv}.
\newblock
\begin{APACrefDOI} \doi{10.48550/ARXIV.2302.08582} \end{APACrefDOI}
\PrintBackRefs{\CurrentBib}

\bibitem [\protect \citeauthoryear {%
Kuhn%
}{%
Kuhn%
}{%
{\protect \APACyear {2022}}%
}]{%
copilotcritic}
\APACinsertmetastar {%
copilotcritic}%
\begin{APACrefauthors}%
Kuhn, B\BPBI M.%
\end{APACrefauthors}%
\unskip\
\newblock
\APACrefYearMonthDay{2022}{}{}.
\newblock
\APACrefbtitle {If Software is My Copilot, Who Programmed My Software?} {If
  software is my copilot, who programmed my software?}
\newblock
\APACaddressPublisher{}{Software Freedom Conservancy}.
\newblock
\begin{APACrefURL}
  \url{https://sfconservancy.org/blog/2022/feb/03/github-copilot-copyleft-gpl/}
  \end{APACrefURL}
\PrintBackRefs{\CurrentBib}

\bibitem [\protect \citeauthoryear {%
Lipton%
}{%
Lipton%
}{%
{\protect \APACyear {2018}}%
}]{%
1606.03490.pdf}
\APACinsertmetastar {%
1606.03490.pdf}%
\begin{APACrefauthors}%
Lipton, Z\BPBI C.%
\end{APACrefauthors}%
\unskip\
\newblock
\APACrefYearMonthDay{2018}{jun}{}.
\newblock
{\BBOQ}\APACrefatitle {The Mythos of Model Interpretability: In Machine
  Learning, the Concept of Interpretability is Both Important and Slippery.}
  {The mythos of model interpretability: In machine learning, the concept of
  interpretability is both important and slippery.}{\BBCQ}
\newblock
\APACjournalVolNumPages{Queue}{16}{3}{31–57}.
\newblock
\begin{APACrefDOI} \doi{10.1145/3236386.3241340} \end{APACrefDOI}
\PrintBackRefs{\CurrentBib}

\bibitem [\protect \citeauthoryear {%
Liu%
\ \BBA {} Chilton%
}{%
Liu%
\ \BBA {} Chilton%
}{%
{\protect \APACyear {2022}}%
}]{%
chilton}
\APACinsertmetastar {%
chilton}%
\begin{APACrefauthors}%
Liu, V.%
\BCBT {}\ \BBA {} Chilton, L\BPBI B.%
\end{APACrefauthors}%
\unskip\
\newblock
\APACrefYearMonthDay{2022}{}{}.
\newblock
{\BBOQ}\APACrefatitle {Design Guidelines for Prompt Engineering Text-to-Image
  Generative Models} {Design guidelines for prompt engineering text-to-image
  generative models}.{\BBCQ}
\newblock
\BIn{} \APACrefbtitle {Proceedings of the 2022 CHI Conference on Human Factors
  in Computing Systems.} {Proceedings of the 2022 chi conference on human
  factors in computing systems.}
\newblock
\APACaddressPublisher{New York, NY, USA}{Association for Computing Machinery}.
\newblock
\begin{APACrefDOI} \doi{10.1145/3491102.3501825} \end{APACrefDOI}
\PrintBackRefs{\CurrentBib}

\bibitem [\protect \citeauthoryear {%
Mansimov%
, Parisotto%
, Ba%
\BCBL {}\ \BBA {} Salakhutdinov%
}{%
Mansimov%
\ \protect \BOthers {.}}{%
{\protect \APACyear {2016}}%
}]{%
1511.02793.pdf}
\APACinsertmetastar {%
1511.02793.pdf}%
\begin{APACrefauthors}%
Mansimov, E.%
, Parisotto, E.%
, Ba, J.%
\BCBL {}\ \BBA {} Salakhutdinov, R.%
\end{APACrefauthors}%
\unskip\
\newblock
\APACrefYearMonthDay{2016}{}{}.
\newblock
{\BBOQ}\APACrefatitle {Generating Images from Captions with Attention}
  {Generating images from captions with attention}.{\BBCQ}
\newblock
\BIn{} \APACrefbtitle {International Conference on Learning Representations.}
  {International conference on learning representations.}
\PrintBackRefs{\CurrentBib}

\bibitem [\protect \citeauthoryear {%
Marche%
}{%
Marche%
}{%
{\protect \APACyear {2022}}%
}]{%
Dont_Fear_the_Artwork_of_the_Future_-_The_Atlantic.pdf}
\APACinsertmetastar {%
Dont_Fear_the_Artwork_of_the_Future_-_The_Atlantic.pdf}%
\begin{APACrefauthors}%
Marche, S.%
\end{APACrefauthors}%
\unskip\
\newblock
\APACrefYearMonthDay{2022}{Sep}{27}.
\newblock
\APACrefbtitle {We're Witnessing the Birth of a New Artistic Medium.} {We're
  witnessing the birth of a new artistic medium.}
\newblock
\APACaddressPublisher{}{The Atlantic}.
\newblock
\begin{APACrefURL}
  \url{https://www.theatlantic.com/technology/archive/2022/09/ai-art-generators-future/671568/}
  \end{APACrefURL}
\PrintBackRefs{\CurrentBib}

\bibitem [\protect \citeauthoryear {%
Matejka%
\ \protect \BOthers {.}}{%
Matejka%
\ \protect \BOthers {.}}{%
{\protect \APACyear {2018}}%
}]{%
DreamLens-CHI-Postrebuttal.pdf}
\APACinsertmetastar {%
DreamLens-CHI-Postrebuttal.pdf}%
\begin{APACrefauthors}%
Matejka, J.%
, Glueck, M.%
, Bradner, E.%
, Hashemi, A.%
, Grossman, T.%
\BCBL {}\ \BBA {} Fitzmaurice, G.%
\end{APACrefauthors}%
\unskip\
\newblock
\APACrefYearMonthDay{2018}{}{}.
\newblock
{\BBOQ}\APACrefatitle {Dream {Lens}: Exploration and Visualization of
  Large-Scale Generative Design Datasets} {Dream {Lens}: Exploration and
  visualization of large-scale generative design datasets}.{\BBCQ}
\newblock
\BIn{} \APACrefbtitle {Proceedings of the 2018 CHI Conference on Human Factors
  in Computing Systems} {Proceedings of the 2018 chi conference on human
  factors in computing systems}\ (\BPG~1–12).
\newblock
\APACaddressPublisher{New York, NY, USA}{Association for Computing Machinery}.
\newblock
\begin{APACrefDOI} \doi{10.1145/3173574.3173943} \end{APACrefDOI}
\PrintBackRefs{\CurrentBib}

\bibitem [\protect \citeauthoryear {%
Mialon%
\ \protect \BOthers {.}}{%
Mialon%
\ \protect \BOthers {.}}{%
{\protect \APACyear {2023}}%
}]{%
2302.07842.pdf}
\APACinsertmetastar {%
2302.07842.pdf}%
\begin{APACrefauthors}%
Mialon, G.%
, Dessì, R.%
, Lomeli, M.%
, Nalmpantis, C.%
, Pasunuru, R.%
, Raileanu, R.%
\BDBL {}Scialom, T.%
\end{APACrefauthors}%
\unskip\
\newblock
\APACrefYearMonthDay{2023}{}{}.
\newblock
\APACrefbtitle {Augmented Language Models: a Survey.} {Augmented language
  models: a survey.}
\newblock
\APACaddressPublisher{}{arXiv}.
\newblock
\begin{APACrefDOI} \doi{10.48550/ARXIV.2302.07842} \end{APACrefDOI}
\PrintBackRefs{\CurrentBib}

\bibitem [\protect \citeauthoryear {%
Milton%
\ \protect \BOthers {.}}{%
Milton%
\ \protect \BOthers {.}}{%
{\protect \APACyear {2021}}%
}]{%
tgab035.pdf}
\APACinsertmetastar {%
tgab035.pdf}%
\begin{APACrefauthors}%
Milton, F.%
, Fulford, J.%
, Dance, C.%
, Gaddum, J.%
, Heuerman-Williamson, B.%
, Jones, K.%
\BDBL {}Zeman, A.%
\end{APACrefauthors}%
\unskip\
\newblock
\APACrefYearMonthDay{2021}{05}{}.
\newblock
{\BBOQ}\APACrefatitle {{Behavioral and Neural Signatures of Visual Imagery
  Vividness Extremes: Aphantasia versus Hyperphantasia}} {{Behavioral and
  Neural Signatures of Visual Imagery Vividness Extremes: Aphantasia versus
  Hyperphantasia}}.{\BBCQ}
\newblock
\APACjournalVolNumPages{Cerebral Cortex Communications}{2}{2}{}.
\newblock
\APACrefnote{tgab035}
\newblock
\begin{APACrefDOI} \doi{10.1093/texcom/tgab035} \end{APACrefDOI}
\PrintBackRefs{\CurrentBib}

\bibitem [\protect \citeauthoryear {%
Mishra%
\ \protect \BOthers {.}}{%
Mishra%
\ \protect \BOthers {.}}{%
{\protect \APACyear {2021}}%
}]{%
2104.08378.pdf}
\APACinsertmetastar {%
2104.08378.pdf}%
\begin{APACrefauthors}%
Mishra, A.%
, Latorre, J\BPBI A.%
, Pool, J.%
, Stosic, D.%
, Stosic, D.%
, Venkatesh, G.%
\BDBL {}Micikevicius, P.%
\end{APACrefauthors}%
\unskip\
\newblock
\APACrefYearMonthDay{2021}{}{}.
\newblock
\APACrefbtitle {Accelerating Sparse Deep Neural Networks.} {Accelerating sparse
  deep neural networks.}
\newblock
\APACaddressPublisher{}{arXiv}.
\newblock
\begin{APACrefDOI} \doi{10.48550/ARXIV.2104.08378} \end{APACrefDOI}
\PrintBackRefs{\CurrentBib}

\bibitem [\protect \citeauthoryear {%
Mok%
}{%
Mok%
}{%
{\protect \APACyear {2023}}%
}]{%
Stack.pdf}
\APACinsertmetastar {%
Stack.pdf}%
\begin{APACrefauthors}%
Mok, K.%
\end{APACrefauthors}%
\unskip\
\newblock
\APACrefYearMonthDay{2023}{}{}.
\newblock
\APACrefbtitle {The Power and Ethical Dilemma of AI Image Generation Models.}
  {The power and ethical dilemma of ai image generation models.}
\newblock
\begin{APACrefURL}
  \url{https://thenewstack.io/the-power-and-ethical-dilemma-of-ai-image-generation-models/}
  \end{APACrefURL}
\PrintBackRefs{\CurrentBib}

\bibitem [\protect \citeauthoryear {%
Monroe%
}{%
Monroe%
}{%
{\protect \APACyear {2021}}%
}]{%
3490155.pdf}
\APACinsertmetastar {%
3490155.pdf}%
\begin{APACrefauthors}%
Monroe, D.%
\end{APACrefauthors}%
\unskip\
\newblock
\APACrefYearMonthDay{2021}{nov}{}.
\newblock
{\BBOQ}\APACrefatitle {Trouble at the Source} {Trouble at the source}.{\BBCQ}
\newblock
\APACjournalVolNumPages{Commun. ACM}{64}{12}{17–19}.
\newblock
\begin{APACrefDOI} \doi{10.1145/3490155} \end{APACrefDOI}
\PrintBackRefs{\CurrentBib}

\bibitem [\protect \citeauthoryear {%
Mountstephens%
\ \BBA {} Teo%
}{%
Mountstephens%
\ \BBA {} Teo%
}{%
{\protect \APACyear {2020}}%
}]{%
computers-09-00080.pdf}
\APACinsertmetastar {%
computers-09-00080.pdf}%
\begin{APACrefauthors}%
Mountstephens, J.%
\BCBT {}\ \BBA {} Teo, J.%
\end{APACrefauthors}%
\unskip\
\newblock
\APACrefYearMonthDay{2020}{}{}.
\newblock
{\BBOQ}\APACrefatitle {Progress and Challenges in Generative Product Design: A
  Review of Systems} {Progress and challenges in generative product design: A
  review of systems}.{\BBCQ}
\newblock
\APACjournalVolNumPages{Computers}{9}{4}{}.
\newblock
\begin{APACrefDOI} \doi{10.3390/computers9040080} \end{APACrefDOI}
\PrintBackRefs{\CurrentBib}

\bibitem [\protect \citeauthoryear {%
Murdock%
\ \BBA {} Wang%
}{%
Murdock%
\ \BBA {} Wang%
}{%
{\protect \APACyear {2021}}%
}]{%
bigsleepgithub}
\APACinsertmetastar {%
bigsleepgithub}%
\begin{APACrefauthors}%
Murdock, R.%
\BCBT {}\ \BBA {} Wang, P.%
\end{APACrefauthors}%
\unskip\
\newblock
\APACrefYearMonthDay{2021}{}{}.
\newblock
\APACrefbtitle {Big Sleep.} {Big sleep.}
\newblock
\begin{APACrefURL} \url{https://github.com/lucidrains/big-sleep}
  \end{APACrefURL}
\PrintBackRefs{\CurrentBib}

\bibitem [\protect \citeauthoryear {%
Nech%
\ \BBA {} Kemelmacher-Shlizerman%
}{%
Nech%
\ \BBA {} Kemelmacher-Shlizerman%
}{%
{\protect \APACyear {2017}}%
}]{%
1705.00393.pdf}
\APACinsertmetastar {%
1705.00393.pdf}%
\begin{APACrefauthors}%
Nech, A.%
\BCBT {}\ \BBA {} Kemelmacher-Shlizerman, I.%
\end{APACrefauthors}%
\unskip\
\newblock
\APACrefYearMonthDay{2017}{}{}.
\newblock
{\BBOQ}\APACrefatitle {Level Playing Field For Million Scale Face Recognition}
  {Level playing field for million scale face recognition}.{\BBCQ}
\newblock
\BIn{} \APACrefbtitle {Proceedings of the IEEE Conference on Computer Vision
  and Pattern Recognition.} {Proceedings of the ieee conference on computer
  vision and pattern recognition.}
\PrintBackRefs{\CurrentBib}

\bibitem [\protect \citeauthoryear {%
Olson%
}{%
Olson%
}{%
{\protect \APACyear {2022}}%
}]{%
messy.pdf}
\APACinsertmetastar {%
messy.pdf}%
\begin{APACrefauthors}%
Olson, P.%
\end{APACrefauthors}%
\unskip\
\newblock
\APACrefYearMonthDay{2022}{}{}.
\newblock
\APACrefbtitle {Creative {AI} Is Generating Some Messy Problems.} {Creative
  {AI} is generating some messy problems.}
\newblock
\APACaddressPublisher{}{Bloomberg}.
\newblock
\begin{APACrefURL}
  \url{https://www.washingtonpost.com/business/creative-ai-is-generating-some-messy-problems/2022/11/28/be2b2efc-6ee2-11ed-867c-8ec695e4afcd_story.html}
  \end{APACrefURL}
\PrintBackRefs{\CurrentBib}

\bibitem [\protect \citeauthoryear {%
Oppenlaender%
}{%
Oppenlaender%
}{%
{\protect \APACyear {2022}}%
{\protect \APACexlab {{\protect \BCnt {1}}}}}]{%
mindtrek-creativity}
\APACinsertmetastar {%
mindtrek-creativity}%
\begin{APACrefauthors}%
Oppenlaender, J.%
\end{APACrefauthors}%
\unskip\
\newblock
\APACrefYearMonthDay{2022{\protect \BCnt {1}}}{}{}.
\newblock
{\BBOQ}\APACrefatitle {The Creativity of Text-to-Image Generation} {The
  creativity of text-to-image generation}.{\BBCQ}
\newblock
\BIn{} \APACrefbtitle {25th {International} {Academic} {Mindtrek} Conference}
  {25th {International} {Academic} {Mindtrek} conference}\ (\BPG~192–202).
\newblock
\APACaddressPublisher{New York, NY, USA}{Association for Computing Machinery}.
\newblock
\begin{APACrefDOI} \doi{10.1145/3569219.3569352} \end{APACrefDOI}
\PrintBackRefs{\CurrentBib}

\bibitem [\protect \citeauthoryear {%
Oppenlaender%
}{%
Oppenlaender%
}{%
{\protect \APACyear {2022}}%
{\protect \APACexlab {{\protect \BCnt {2}}}}}]{%
taxonomy}
\APACinsertmetastar {%
taxonomy}%
\begin{APACrefauthors}%
Oppenlaender, J.%
\end{APACrefauthors}%
\unskip\
\newblock
\APACrefYearMonthDay{2022{\protect \BCnt {2}}}{}{}.
\newblock
\APACrefbtitle {A Taxonomy of Prompt Modifiers for Text-to-Image Generation.}
  {A taxonomy of prompt modifiers for text-to-image generation.}
\newblock
\APACaddressPublisher{}{arXiv}.
\newblock
\begin{APACrefDOI} \doi{10.48550/ARXIV.2204.13988} \end{APACrefDOI}
\PrintBackRefs{\CurrentBib}

\bibitem [\protect \citeauthoryear {%
Oppenlaender%
, Silvennoinen%
, Paananen%
\BCBL {}\ \BBA {} Visuri%
}{%
Oppenlaender%
\ \protect \BOthers {.}}{%
{\protect \APACyear {2023}}%
}]{%
oppenlaender2023perceptions}
\APACinsertmetastar {%
oppenlaender2023perceptions}%
\begin{APACrefauthors}%
Oppenlaender, J.%
, Silvennoinen, J.%
, Paananen, V.%
\BCBL {}\ \BBA {} Visuri, A.%
\end{APACrefauthors}%
\unskip\
\newblock
\APACrefYearMonthDay{2023}{}{}.
\newblock
\APACrefbtitle {Perceptions and Realities of Text-to-Image Generation.}
  {Perceptions and realities of text-to-image generation.}
\newblock
\begin{APACrefDOI} \doi{10.48550/arXiv.2306.08363} \end{APACrefDOI}
\PrintBackRefs{\CurrentBib}

\bibitem [\protect \citeauthoryear {%
Paananen%
, Oppenlaender%
\BCBL {}\ \BBA {} Visuri%
}{%
Paananen%
\ \protect \BOthers {.}}{%
{\protect \APACyear {2023}}%
}]{%
paananen2023architecture}
\APACinsertmetastar {%
paananen2023architecture}%
\begin{APACrefauthors}%
Paananen, V.%
, Oppenlaender, J.%
\BCBL {}\ \BBA {} Visuri, A.%
\end{APACrefauthors}%
\unskip\
\newblock
\APACrefYearMonthDay{2023}{}{}.
\newblock
\APACrefbtitle {Using Text-to-Image Generation for Architectural Design
  Ideation.} {Using text-to-image generation for architectural design
  ideation.}
\newblock
\begin{APACrefDOI} \doi{10.48550/arXiv.2304.10182} \end{APACrefDOI}
\PrintBackRefs{\CurrentBib}

\bibitem [\protect \citeauthoryear {%
Parsons%
}{%
Parsons%
}{%
{\protect \APACyear {2022}}%
}]{%
dallepromptbook}
\APACinsertmetastar {%
dallepromptbook}%
\begin{APACrefauthors}%
Parsons, G.%
\end{APACrefauthors}%
\unskip\
\newblock
\APACrefYearMonthDay{2022}{}{}.
\newblock
\APACrefbtitle {The {DALL·E} 2 {Prompt} {Book}.} {The {DALL·E} 2 {Prompt}
  {Book}.}
\newblock
\begin{APACrefURL} \url{https://dallery.gallery/the-dalle-2-prompt-book/}
  \end{APACrefURL}
\PrintBackRefs{\CurrentBib}

\bibitem [\protect \citeauthoryear {%
Perez%
\ \protect \BOthers {.}}{%
Perez%
\ \protect \BOthers {.}}{%
{\protect \APACyear {2022}}%
}]{%
model-written-evals.pdf}
\APACinsertmetastar {%
model-written-evals.pdf}%
\begin{APACrefauthors}%
Perez, E.%
, Ringer, S.%
, Lukošiūtė, K.%
, Nguyen, K.%
, Chen, E.%
, Heiner, S.%
\BDBL {}Kaplan, J.%
\end{APACrefauthors}%
\unskip\
\newblock
\APACrefYearMonthDay{2022}{}{}.
\newblock
\APACrefbtitle {Discovering Language Model Behaviors with Model-Written
  Evaluations.} {Discovering language model behaviors with model-written
  evaluations.}
\newblock
\APACaddressPublisher{}{arXiv}.
\newblock
\begin{APACrefDOI} \doi{10.48550/ARXIV.2212.09251} \end{APACrefDOI}
\PrintBackRefs{\CurrentBib}

\bibitem [\protect \citeauthoryear {%
Poursabzi-Sangdeh%
, Goldstein%
, Hofman%
, Wortman~Vaughan%
\BCBL {}\ \BBA {} Wallach%
}{%
Poursabzi-Sangdeh%
\ \protect \BOthers {.}}{%
{\protect \APACyear {2021}}%
}]{%
1802.07810.pdf}
\APACinsertmetastar {%
1802.07810.pdf}%
\begin{APACrefauthors}%
Poursabzi-Sangdeh, F.%
, Goldstein, D\BPBI G.%
, Hofman, J\BPBI M.%
, Wortman~Vaughan, J\BPBI W.%
\BCBL {}\ \BBA {} Wallach, H.%
\end{APACrefauthors}%
\unskip\
\newblock
\APACrefYearMonthDay{2021}{}{}.
\newblock
{\BBOQ}\APACrefatitle {Manipulating and Measuring Model Interpretability}
  {Manipulating and measuring model interpretability}.{\BBCQ}
\newblock
\BIn{} \APACrefbtitle {Proceedings of the 2021 CHI Conference on Human Factors
  in Computing Systems.} {Proceedings of the 2021 chi conference on human
  factors in computing systems.}
\newblock
\APACaddressPublisher{New York, NY, USA}{Association for Computing Machinery}.
\newblock
\begin{APACrefDOI} \doi{10.1145/3411764.3445315} \end{APACrefDOI}
\PrintBackRefs{\CurrentBib}

\bibitem [\protect \citeauthoryear {%
Qu%
\ \protect \BOthers {.}}{%
Qu%
\ \protect \BOthers {.}}{%
{\protect \APACyear {2023}}%
}]{%
2305.13873.pdf}
\APACinsertmetastar {%
2305.13873.pdf}%
\begin{APACrefauthors}%
Qu, Y.%
, Shen, X.%
, He, X.%
, Backes, M.%
, Zannettou, S.%
\BCBL {}\ \BBA {} Zhang, Y.%
\end{APACrefauthors}%
\unskip\
\newblock
\APACrefYearMonthDay{2023}{}{}.
\newblock
\APACrefbtitle {Unsafe Diffusion: On the Generation of Unsafe Images and
  Hateful Memes From Text-To-Image Models.} {Unsafe diffusion: On the
  generation of unsafe images and hateful memes from text-to-image models.}
\newblock
\APACaddressPublisher{New York, NY, USA}{Association for Computing Machinery}.
\PrintBackRefs{\CurrentBib}

\bibitem [\protect \citeauthoryear {%
Radford%
\ \protect \BOthers {.}}{%
Radford%
\ \protect \BOthers {.}}{%
{\protect \APACyear {2021}}%
}]{%
CLIP}
\APACinsertmetastar {%
CLIP}%
\begin{APACrefauthors}%
Radford, A.%
, Kim, J\BPBI W.%
, Hallacy, C.%
, Ramesh, A.%
, Goh, G.%
, Agarwal, S.%
\BDBL {}Sutskever, I.%
\end{APACrefauthors}%
\unskip\
\newblock
\APACrefYearMonthDay{2021}{}{}.
\newblock
{\BBOQ}\APACrefatitle {Learning Transferable Visual Models From Natural
  Language Supervision} {Learning transferable visual models from natural
  language supervision}.{\BBCQ}
\newblock
\BIn{} M.~Meila\ \BBA {} T.~Zhang\ (\BEDS), \APACrefbtitle {Proceedings of the
  38th International Conference on Machine Learning} {Proceedings of the 38th
  international conference on machine learning}\ (\BVOL~139, \BPGS\
  8748--8763).
\newblock
\APACaddressPublisher{}{PMLR}.
\PrintBackRefs{\CurrentBib}

\bibitem [\protect \citeauthoryear {%
Radford%
\ \protect \BOthers {.}}{%
Radford%
\ \protect \BOthers {.}}{%
{\protect \APACyear {2019}}%
}]{%
language_models_are_unsupervised_multitask_learners.pdf}
\APACinsertmetastar {%
language_models_are_unsupervised_multitask_learners.pdf}%
\begin{APACrefauthors}%
Radford, A.%
, Wu, J.%
, Child, R.%
, Luan, D.%
, Amodei, D.%
\BCBL {}\ \BBA {} Sutskever, I.%
\end{APACrefauthors}%
\unskip\
\newblock
\APACrefYearMonthDay{2019}{}{}.
\newblock
{\BBOQ}\APACrefatitle {Language Models are Unsupervised Multitask Learners}
  {Language models are unsupervised multitask learners}.{\BBCQ}
\newblock
\APACjournalVolNumPages{{OpenAI} blog}{1}{8}{9}.
\PrintBackRefs{\CurrentBib}

\bibitem [\protect \citeauthoryear {%
Raffel%
\ \protect \BOthers {.}}{%
Raffel%
\ \protect \BOthers {.}}{%
{\protect \APACyear {2022}}%
}]{%
1910.10683.pdf}
\APACinsertmetastar {%
1910.10683.pdf}%
\begin{APACrefauthors}%
Raffel, C.%
, Shazeer, N.%
, Roberts, A.%
, Lee, K.%
, Narang, S.%
, Matena, M.%
\BDBL {}Liu, P\BPBI J.%
\end{APACrefauthors}%
\unskip\
\newblock
\APACrefYearMonthDay{2022}{jun}{}.
\newblock
{\BBOQ}\APACrefatitle {Exploring the Limits of Transfer Learning with a Unified
  Text-to-Text Transformer} {Exploring the limits of transfer learning with a
  unified text-to-text transformer}.{\BBCQ}
\newblock
\APACjournalVolNumPages{J. Mach. Learn. Res.}{21}{1}{}.
\PrintBackRefs{\CurrentBib}

\bibitem [\protect \citeauthoryear {%
Ramesh%
, Dhariwal%
, Nichol%
, Chu%
\BCBL {}\ \BBA {} Chen%
}{%
Ramesh%
\ \protect \BOthers {.}}{%
{\protect \APACyear {2022}}%
}]{%
dalle2}
\APACinsertmetastar {%
dalle2}%
\begin{APACrefauthors}%
Ramesh, A.%
, Dhariwal, P.%
, Nichol, A.%
, Chu, C.%
\BCBL {}\ \BBA {} Chen, M.%
\end{APACrefauthors}%
\unskip\
\newblock
\APACrefYearMonthDay{2022}{}{}.
\newblock
\APACrefbtitle {Hierarchical Text-Conditional Image Generation with {CLIP}
  Latents.} {Hierarchical text-conditional image generation with {CLIP}
  latents.}
\newblock
\APACaddressPublisher{}{arXiv}.
\newblock
\begin{APACrefDOI} \doi{10.48550/ARXIV.2204.06125} \end{APACrefDOI}
\PrintBackRefs{\CurrentBib}

\bibitem [\protect \citeauthoryear {%
Rassin%
, Ravfogel%
\BCBL {}\ \BBA {} Goldberg%
}{%
Rassin%
\ \protect \BOthers {.}}{%
{\protect \APACyear {2022}}%
}]{%
2210.10606.pdf}
\APACinsertmetastar {%
2210.10606.pdf}%
\begin{APACrefauthors}%
Rassin, R.%
, Ravfogel, S.%
\BCBL {}\ \BBA {} Goldberg, Y.%
\end{APACrefauthors}%
\unskip\
\newblock
\APACrefYearMonthDay{2022}{{\APACmonth{12}}}{}.
\newblock
{\BBOQ}\APACrefatitle {{DALLE}-2 is Seeing Double: Flaws in Word-to-Concept
  Mapping in {T}ext2{I}mage Models} {{DALLE}-2 is seeing double: Flaws in
  word-to-concept mapping in {T}ext2{I}mage models}.{\BBCQ}
\newblock
\BIn{} \APACrefbtitle {Proceedings of the Fifth BlackboxNLP Workshop on
  Analyzing and Interpreting Neural Networks for NLP} {Proceedings of the fifth
  blackboxnlp workshop on analyzing and interpreting neural networks for nlp}\
  (\BPGS\ 335--345).
\newblock
\APACaddressPublisher{Abu Dhabi, United Arab Emirates (Hybrid)}{Association for
  Computational Linguistics}.
\newblock
\begin{APACrefURL} \url{https://aclanthology.org/2022.blackboxnlp-1.28}
  \end{APACrefURL}
\PrintBackRefs{\CurrentBib}

\bibitem [\protect \citeauthoryear {%
Reed%
\ \protect \BOthers {.}}{%
Reed%
\ \protect \BOthers {.}}{%
{\protect \APACyear {2016}}%
}]{%
1605.05396.pdf}
\APACinsertmetastar {%
1605.05396.pdf}%
\begin{APACrefauthors}%
Reed, S.%
, Akata, Z.%
, Yan, X.%
, Logeswaran, L.%
, Schiele, B.%
\BCBL {}\ \BBA {} Lee, H.%
\end{APACrefauthors}%
\unskip\
\newblock
\APACrefYearMonthDay{2016}{}{}.
\newblock
\APACrefbtitle {Generative Adversarial Text to Image Synthesis.} {Generative
  adversarial text to image synthesis.}
\newblock
\APACaddressPublisher{}{arXiv}.
\newblock
\begin{APACrefDOI} \doi{10.48550/ARXIV.1605.05396} \end{APACrefDOI}
\PrintBackRefs{\CurrentBib}

\bibitem [\protect \citeauthoryear {%
Rombach%
, Blattmann%
, Lorenz%
, Esser%
\BCBL {}\ \BBA {} Ommer%
}{%
Rombach%
\ \protect \BOthers {.}}{%
{\protect \APACyear {2021}}%
}]{%
latent-diffusion}
\APACinsertmetastar {%
latent-diffusion}%
\begin{APACrefauthors}%
Rombach, R.%
, Blattmann, A.%
, Lorenz, D.%
, Esser, P.%
\BCBL {}\ \BBA {} Ommer, B.%
\end{APACrefauthors}%
\unskip\
\newblock
\APACrefYearMonthDay{2021}{}{}.
\newblock
\APACrefbtitle {High-Resolution Image Synthesis with Latent Diffusion Models.}
  {High-resolution image synthesis with latent diffusion models.}
\PrintBackRefs{\CurrentBib}

\bibitem [\protect \citeauthoryear {%
Rombach%
, Blattmann%
\BCBL {}\ \BBA {} Ommer%
}{%
Rombach%
\ \protect \BOthers {.}}{%
{\protect \APACyear {2022}}%
}]{%
2207.13038.pdf}
\APACinsertmetastar {%
2207.13038.pdf}%
\begin{APACrefauthors}%
Rombach, R.%
, Blattmann, A.%
\BCBL {}\ \BBA {} Ommer, B.%
\end{APACrefauthors}%
\unskip\
\newblock
\APACrefYearMonthDay{2022}{}{}.
\newblock
\APACrefbtitle {Text-Guided Synthesis of Artistic Images with
  Retrieval-Augmented Diffusion Models.} {Text-guided synthesis of artistic
  images with retrieval-augmented diffusion models.}
\newblock
\APACaddressPublisher{}{arXiv}.
\newblock
\begin{APACrefDOI} \doi{10.48550/ARXIV.2207.13038} \end{APACrefDOI}
\PrintBackRefs{\CurrentBib}

\bibitem [\protect \citeauthoryear {%
Saini%
}{%
Saini%
}{%
{\protect \APACyear {2022}}%
}]{%
MisterRuffian}
\APACinsertmetastar {%
MisterRuffian}%
\begin{APACrefauthors}%
Saini, L.%
\end{APACrefauthors}%
\unskip\
\newblock
\APACrefYearMonthDay{2022}{}{}.
\newblock
\APACrefbtitle {MisterRuffian's Latent Artist \& Modifier Encyclopedia.}
  {Misterruffian's latent artist \& modifier encyclopedia.}
\newblock
\begin{APACrefURL}
  \url{https://docs.google.com/spreadsheets/d/1_jgQ9SyvUaBNP1mHHEzZ6HhL_Es1KwBKQtnpnmWW82I}
  \end{APACrefURL}
\PrintBackRefs{\CurrentBib}

\bibitem [\protect \citeauthoryear {%
Salminen%
, Jung%
, Chowdhury%
\BCBL {}\ \BBA {} Jansen%
}{%
Salminen%
\ \protect \BOthers {.}}{%
{\protect \APACyear {2020}}%
}]{%
3334480.3382791.pdf}
\APACinsertmetastar {%
3334480.3382791.pdf}%
\begin{APACrefauthors}%
Salminen, J.%
, Jung, S\BHBI g.%
, Chowdhury, S.%
\BCBL {}\ \BBA {} Jansen, B\BPBI J.%
\end{APACrefauthors}%
\unskip\
\newblock
\APACrefYearMonthDay{2020}{}{}.
\newblock
{\BBOQ}\APACrefatitle {Analyzing Demographic Bias in Artificially Generated
  Facial Pictures} {Analyzing demographic bias in artificially generated facial
  pictures}.{\BBCQ}
\newblock
\BIn{} \APACrefbtitle {Extended Abstracts of the 2020 CHI Conference on Human
  Factors in Computing Systems} {Extended abstracts of the 2020 chi conference
  on human factors in computing systems}\ (\BPG~1–8).
\newblock
\APACaddressPublisher{New York, NY, USA}{Association for Computing Machinery}.
\newblock
\begin{APACrefDOI} \doi{10.1145/3334480.3382791} \end{APACrefDOI}
\PrintBackRefs{\CurrentBib}

\bibitem [\protect \citeauthoryear {%
Schuhmann%
\ \protect \BOthers {.}}{%
Schuhmann%
\ \protect \BOthers {.}}{%
{\protect \APACyear {2022}}%
}]{%
2210.08402.pdf}
\APACinsertmetastar {%
2210.08402.pdf}%
\begin{APACrefauthors}%
Schuhmann, C.%
, Beaumont, R.%
, Vencu, R.%
, Gordon, C\BPBI W.%
, Wightman, R.%
, Cherti, M.%
\BDBL {}Jitsev, J.%
\end{APACrefauthors}%
\unskip\
\newblock
\APACrefYearMonthDay{2022}{}{}.
\newblock
{\BBOQ}\APACrefatitle {{LAION}-{5B}: An open large-scale dataset for training
  next generation image-text models} {{LAION}-{5B}: An open large-scale dataset
  for training next generation image-text models}.{\BBCQ}
\newblock
\BIn{} \APACrefbtitle {Thirty-sixth Conference on Neural Information Processing
  Systems Datasets and Benchmarks Track.} {Thirty-sixth conference on neural
  information processing systems datasets and benchmarks track.}
\PrintBackRefs{\CurrentBib}

\bibitem [\protect \citeauthoryear {%
Schuhmann%
\ \protect \BOthers {.}}{%
Schuhmann%
\ \protect \BOthers {.}}{%
{\protect \APACyear {2021}}%
}]{%
2111.02114.pdf}
\APACinsertmetastar {%
2111.02114.pdf}%
\begin{APACrefauthors}%
Schuhmann, C.%
, Vencu, R.%
, Beaumont, R.%
, Kaczmarczyk, R.%
, Mullis, C.%
, Katta, A.%
\BDBL {}Komatsuzaki, A.%
\end{APACrefauthors}%
\unskip\
\newblock
\APACrefYearMonthDay{2021}{}{}.
\newblock
\APACrefbtitle {{LAION}-{400M}: Open Dataset of {CLIP}-Filtered 400 Million
  Image-Text Pairs.} {{LAION}-{400M}: Open dataset of {CLIP}-filtered 400
  million image-text pairs.}
\newblock
\APACaddressPublisher{}{arXiv}.
\newblock
\begin{APACrefDOI} \doi{10.48550/ARXIV.2111.02114} \end{APACrefDOI}
\PrintBackRefs{\CurrentBib}

\bibitem [\protect \citeauthoryear {%
Shen%
, DeVos%
, Eslami%
\BCBL {}\ \BBA {} Holstein%
}{%
Shen%
\ \protect \BOthers {.}}{%
{\protect \APACyear {2021}}%
}]{%
2105.02980.pdf}
\APACinsertmetastar {%
2105.02980.pdf}%
\begin{APACrefauthors}%
Shen, H.%
, DeVos, A.%
, Eslami, M.%
\BCBL {}\ \BBA {} Holstein, K.%
\end{APACrefauthors}%
\unskip\
\newblock
\APACrefYearMonthDay{2021}{oct}{}.
\newblock
{\BBOQ}\APACrefatitle {Everyday Algorithm Auditing: Understanding the Power of
  Everyday Users in Surfacing Harmful Algorithmic Behaviors} {Everyday
  algorithm auditing: Understanding the power of everyday users in surfacing
  harmful algorithmic behaviors}.{\BBCQ}
\newblock
\APACjournalVolNumPages{Proc. ACM Hum.-Comput. Interact.}{5}{CSCW2}{}.
\newblock
\begin{APACrefDOI} \doi{10.1145/3479577} \end{APACrefDOI}
\PrintBackRefs{\CurrentBib}

\bibitem [\protect \citeauthoryear {%
Shumailov%
\ \protect \BOthers {.}}{%
Shumailov%
\ \protect \BOthers {.}}{%
{\protect \APACyear {2023}}%
}]{%
2305.17493.pdf}
\APACinsertmetastar {%
2305.17493.pdf}%
\begin{APACrefauthors}%
Shumailov, I.%
, Shumaylov, Z.%
, Zhao, Y.%
, Gal, Y.%
, Papernot, N.%
\BCBL {}\ \BBA {} Anderson, R.%
\end{APACrefauthors}%
\unskip\
\newblock
\APACrefYearMonthDay{2023}{}{}.
\newblock
\APACrefbtitle {The Curse of Recursion: Training on Generated Data Makes Models
  Forget.} {The curse of recursion: Training on generated data makes models
  forget.}
\PrintBackRefs{\CurrentBib}

\bibitem [\protect \citeauthoryear {%
Somepalli%
, Singla%
, Goldblum%
, Geiping%
\BCBL {}\ \BBA {} Goldstein%
}{%
Somepalli%
\ \protect \BOthers {.}}{%
{\protect \APACyear {2022}}%
}]{%
2212.03860.pdf}
\APACinsertmetastar {%
2212.03860.pdf}%
\begin{APACrefauthors}%
Somepalli, G.%
, Singla, V.%
, Goldblum, M.%
, Geiping, J.%
\BCBL {}\ \BBA {} Goldstein, T.%
\end{APACrefauthors}%
\unskip\
\newblock
\APACrefYearMonthDay{2022}{}{}.
\newblock
\APACrefbtitle {Diffusion Art or Digital Forgery? {Investigating} Data
  Replication in Diffusion Models.} {Diffusion art or digital forgery?
  {Investigating} data replication in diffusion models.}
\newblock
\APACaddressPublisher{}{arXiv}.
\newblock
\begin{APACrefDOI} \doi{10.48550/ARXIV.2212.03860} \end{APACrefDOI}
\PrintBackRefs{\CurrentBib}

\bibitem [\protect \citeauthoryear {%
{Stability AI}%
}{%
{Stability AI}%
}{%
{\protect \APACyear {2022}}%
}]{%
photoshop2}
\APACinsertmetastar {%
photoshop2}%
\begin{APACrefauthors}%
{Stability AI}.%
\end{APACrefauthors}%
\unskip\
\newblock
\APACrefYearMonthDay{2022}{}{}.
\newblock
\APACrefbtitle {Stability {Photoshop} plugin.} {Stability {Photoshop} plugin.}
\newblock
\begin{APACrefURL}
  \url{https://exchange.adobe.com/apps/cc/114117da/stable-diffusion}
  \end{APACrefURL}
\PrintBackRefs{\CurrentBib}

\bibitem [\protect \citeauthoryear {%
Stackoverflow.com%
}{%
Stackoverflow.com%
}{%
{\protect \APACyear {2022}}%
}]{%
stackoverflow}
\APACinsertmetastar {%
stackoverflow}%
\begin{APACrefauthors}%
Stackoverflow.com.%
\end{APACrefauthors}%
\unskip\
\newblock
\APACrefYearMonthDay{2022}{}{}.
\newblock
\APACrefbtitle {Temporary policy: ChatGPT is banned.} {Temporary policy:
  Chatgpt is banned.}
\newblock
\begin{APACrefURL}
  \url{https://meta.stackoverflow.com/questions/421831/temporary-policy-chatgpt-is-banned}
  \end{APACrefURL}
\PrintBackRefs{\CurrentBib}

\bibitem [\protect \citeauthoryear {%
Toews%
}{%
Toews%
}{%
{\protect \APACyear {2022}}%
}]{%
Predictions}
\APACinsertmetastar {%
Predictions}%
\begin{APACrefauthors}%
Toews, R.%
\end{APACrefauthors}%
\unskip\
\newblock
\APACrefYearMonthDay{2022}{}{}.
\newblock
\APACrefbtitle {4 Predictions About The Wild New World Of Text-To-Image AI.} {4
  predictions about the wild new world of text-to-image ai.}
\newblock
\APACaddressPublisher{}{Forbes}.
\newblock
\begin{APACrefURL}
  \url{https://www.forbes.com/sites/robtoews/2022/09/11/4-hot-takes-about-the-wild-new-world-of-generative-ai/}
  \end{APACrefURL}
\PrintBackRefs{\CurrentBib}

\bibitem [\protect \citeauthoryear {%
Townsend%
}{%
Townsend%
}{%
{\protect \APACyear {2023}}%
}]{%
corecore}
\APACinsertmetastar {%
corecore}%
\begin{APACrefauthors}%
Townsend, C.%
\end{APACrefauthors}%
\unskip\
\newblock
\APACrefYearMonthDay{2023}{}{}.
\newblock
\APACrefbtitle {Explaining corecore: How TikTok's newest trend may be a genuine
  Gen-Z art form.} {Explaining corecore: How tiktok's newest trend may be a
  genuine gen-z art form.}
\newblock
\begin{APACrefURL}
  \url{https://mashable.com/article/explaining-corecore-tiktok}
  \end{APACrefURL}
\PrintBackRefs{\CurrentBib}

\bibitem [\protect \citeauthoryear {%
van~der Wal%
, Jumelet%
, Schulz%
\BCBL {}\ \BBA {} Zuidema%
}{%
van~der Wal%
\ \protect \BOthers {.}}{%
{\protect \APACyear {2022}}%
}]{%
2207.10245.pdf}
\APACinsertmetastar {%
2207.10245.pdf}%
\begin{APACrefauthors}%
van~der Wal, O.%
, Jumelet, J.%
, Schulz, K.%
\BCBL {}\ \BBA {} Zuidema, W.%
\end{APACrefauthors}%
\unskip\
\newblock
\APACrefYearMonthDay{2022}{}{}.
\newblock
\APACrefbtitle {The Birth of Bias: A case study on the evolution of gender bias
  in an English language model.} {The birth of bias: A case study on the
  evolution of gender bias in an english language model.}
\newblock
\APACaddressPublisher{}{arXiv}.
\newblock
\begin{APACrefDOI} \doi{10.48550/ARXIV.2207.10245} \end{APACrefDOI}
\PrintBackRefs{\CurrentBib}

\bibitem [\protect \citeauthoryear {%
Vincent%
}{%
Vincent%
}{%
{\protect \APACyear {{\protect \bibnodate {}}}}%
}]{%
getty-infringement}
\APACinsertmetastar {%
getty-infringement}%
\begin{APACrefauthors}%
Vincent, J.%
\end{APACrefauthors}%
\unskip\
\newblock
\APACrefYearMonthDay{{\protect \bibnodate {}}}{}{}.
\newblock
\APACrefbtitle {Getty Images sues AI art generator Stable Diffusion in the US
  for copyright infringement.} {Getty images sues ai art generator stable
  diffusion in the us for copyright infringement.}
\newblock
\APACaddressPublisher{}{The Verge}.
\newblock
\begin{APACrefURL}
  \url{https://www.theverge.com/2023/2/6/23587393/ai-art-copyright-lawsuit-getty-images-stable-diffusion}
  \end{APACrefURL}
\PrintBackRefs{\CurrentBib}

\bibitem [\protect \citeauthoryear {%
Vincent%
}{%
Vincent%
}{%
{\protect \APACyear {2022}}%
{\protect \APACexlab {{\protect \BCnt {1}}}}}]{%
getty}
\APACinsertmetastar {%
getty}%
\begin{APACrefauthors}%
Vincent, J.%
\end{APACrefauthors}%
\unskip\
\newblock
\APACrefYearMonthDay{2022{\protect \BCnt {1}}}{}{}.
\newblock
\APACrefbtitle {Getty Images bans AI-generated content over fears of legal
  challenges.} {Getty images bans ai-generated content over fears of legal
  challenges.}
\newblock
\APACaddressPublisher{}{The Verge}.
\newblock
\begin{APACrefURL}
  \url{https://www.theverge.com/2022/9/21/23364696/getty-images-ai-ban-generated-artwork-illustration-copyright}
  \end{APACrefURL}
\PrintBackRefs{\CurrentBib}

\bibitem [\protect \citeauthoryear {%
Vincent%
}{%
Vincent%
}{%
{\protect \APACyear {2022}}%
{\protect \APACexlab {{\protect \BCnt {2}}}}}]{%
Verge.pdf}
\APACinsertmetastar {%
Verge.pdf}%
\begin{APACrefauthors}%
Vincent, J.%
\end{APACrefauthors}%
\unskip\
\newblock
\APACrefYearMonthDay{2022{\protect \BCnt {2}}}{}{}.
\newblock
\APACrefbtitle {The lawsuit that could rewrite the rules of AI copyright.} {The
  lawsuit that could rewrite the rules of ai copyright.}
\newblock
\APACaddressPublisher{}{The Verge}.
\newblock
\begin{APACrefURL}
  \url{https://www.theverge.com/2022/11/8/23446821/microsoft-openai-github-copilot-class-action-lawsuit-ai-copyright-violation-training-data}
  \end{APACrefURL}
\PrintBackRefs{\CurrentBib}

\bibitem [\protect \citeauthoryear {%
Wang%
}{%
Wang%
}{%
{\protect \APACyear {2022}}%
}]{%
pehype}
\APACinsertmetastar {%
pehype}%
\begin{APACrefauthors}%
Wang, S.%
\end{APACrefauthors}%
\unskip\
\newblock
\APACrefYearMonthDay{2022}{}{}.
\newblock
\APACrefbtitle {Why ``Prompt Engineering'' and ``Generative {AI}'' are
  overhyped.} {Why ``prompt engineering'' and ``generative {AI}'' are
  overhyped.}
\newblock
\begin{APACrefURL}
  \url{https://lspace.swyx.io/p/why-prompt-engineering-and-generative}
  \end{APACrefURL}
\PrintBackRefs{\CurrentBib}

\bibitem [\protect \citeauthoryear {%
Williams%
, Brooks%
\BCBL {}\ \BBA {} Shmargad%
}{%
Williams%
\ \protect \BOthers {.}}{%
{\protect \APACyear {2018}}%
}]{%
jinfopoli_8_1_78.pdf}
\APACinsertmetastar {%
jinfopoli_8_1_78.pdf}%
\begin{APACrefauthors}%
Williams, B\BPBI A.%
, Brooks, C\BPBI F.%
\BCBL {}\ \BBA {} Shmargad, Y.%
\end{APACrefauthors}%
\unskip\
\newblock
\APACrefYearMonthDay{2018}{03}{}.
\newblock
{\BBOQ}\APACrefatitle {{How Algorithms Discriminate Based on Data They Lack:
  Challenges, Solutions, and Policy Implications}} {{How Algorithms
  Discriminate Based on Data They Lack: Challenges, Solutions, and Policy
  Implications}}.{\BBCQ}
\newblock
\APACjournalVolNumPages{Journal of Information Policy}{8}{}{78-115}.
\newblock
\begin{APACrefDOI} \doi{10.5325/jinfopoli.8.2018.0078} \end{APACrefDOI}
\PrintBackRefs{\CurrentBib}

\bibitem [\protect \citeauthoryear {%
Wilmer%
, Sherman%
\BCBL {}\ \BBA {} Chein%
}{%
Wilmer%
\ \protect \BOthers {.}}{%
{\protect \APACyear {2017}}%
}]{%
fpsyg-08-00605.pdf}
\APACinsertmetastar {%
fpsyg-08-00605.pdf}%
\begin{APACrefauthors}%
Wilmer, H\BPBI H.%
, Sherman, L\BPBI E.%
\BCBL {}\ \BBA {} Chein, J\BPBI M.%
\end{APACrefauthors}%
\unskip\
\newblock
\APACrefYearMonthDay{2017}{}{}.
\newblock
{\BBOQ}\APACrefatitle {Smartphones and Cognition: A Review of Research
  Exploring the Links between Mobile Technology Habits and Cognitive
  Functioning} {Smartphones and cognition: A review of research exploring the
  links between mobile technology habits and cognitive functioning}.{\BBCQ}
\newblock
\APACjournalVolNumPages{Frontiers in Psychology}{8}{}{}.
\newblock
\begin{APACrefDOI} \doi{10.3389/fpsyg.2017.00605} \end{APACrefDOI}
\PrintBackRefs{\CurrentBib}

\bibitem [\protect \citeauthoryear {%
Zammit%
, Liapis%
\BCBL {}\ \BBA {} Yannakakis%
}{%
Zammit%
\ \protect \BOthers {.}}{%
{\protect \APACyear {2022}}%
}]{%
ICCC-2022_15L_Zammit-et-al..pdf}
\APACinsertmetastar {%
ICCC-2022_15L_Zammit-et-al..pdf}%
\begin{APACrefauthors}%
Zammit, M.%
, Liapis, A.%
\BCBL {}\ \BBA {} Yannakakis, G\BPBI N.%
\end{APACrefauthors}%
\unskip\
\newblock
\APACrefYearMonthDay{2022}{}{}.
\newblock
\APACrefbtitle {Seeding Diversity into AI Art.} {Seeding diversity into ai
  art.}
\newblock
\APACaddressPublisher{}{arXiv}.
\newblock
\begin{APACrefDOI} \doi{10.48550/ARXIV.2205.00804} \end{APACrefDOI}
\PrintBackRefs{\CurrentBib}

\bibitem [\protect \citeauthoryear {%
Zeman%
\ \protect \BOthers {.}}{%
Zeman%
\ \protect \BOthers {.}}{%
{\protect \APACyear {2020}}%
}]{%
1-s2.0-S0010945220301404-main.pdf}
\APACinsertmetastar {%
1-s2.0-S0010945220301404-main.pdf}%
\begin{APACrefauthors}%
Zeman, A.%
, Milton, F.%
, {Della Sala}, S.%
, Dewar, M.%
, Frayling, T.%
, Gaddum, J.%
\BDBL {}Winlove, C.%
\end{APACrefauthors}%
\unskip\
\newblock
\APACrefYearMonthDay{2020}{}{}.
\newblock
{\BBOQ}\APACrefatitle {Phantasia -- {The} psychological significance of
  lifelong visual imagery vividness extremes} {Phantasia -- {The} psychological
  significance of lifelong visual imagery vividness extremes}.{\BBCQ}
\newblock
\APACjournalVolNumPages{Cortex}{130}{}{426-440}.
\newblock
\begin{APACrefDOI} \doi{10.1016/j.cortex.2020.04.003} \end{APACrefDOI}
\PrintBackRefs{\CurrentBib}

\end{thebibliography}


\end{document}